\documentclass[11pt]{article}
\usepackage{amsmath,amsfonts,amssymb}
\usepackage{graphicx}
\usepackage{epsfig,epsf}
\usepackage[colorlinks=true,
            linkcolor=red,
            urlcolor=blue,
            citecolor=blue]{hyperref}
\def\bar {\overline}

\def\be {\begin{equation}}
\def\ee {\end{equation}}
\def\beq {\begin{equation}}
\def\eeq {\end{equation}}
\def\bea {\begin{eqnarray}}
\def\eea {\end{eqnarray}}

\def\bra {\langle}
\def\ket {\rangle}

\def\beq{\begin{equation}}
\def\eeq{\end{equation}}
\def\barr{\begin{array}}
\def\earr{\end{array}}

\def\opcit(#1){ {\em op. cit.}, #1}

\def\issue(#1,#2,#3){#1, #2 (#3)} 

\def\APP(#1,#2,#3){Acta Phys.\ Polon.\ \issue(#1,#2,#3)}
\def\ARNPS(#1,#2,#3){Ann.\ Rev.\ Nucl.\ Part.\ Sci.\ \issue(#1,#2,#3)}
\def\CPC(#1,#2,#3){Comp.\ Phys.\ Comm.\ \issue(#1,#2,#3)}
\def\CIP(#1,#2,#3){Comput.\ Phys.\ \issue(#1,#2,#3)}
\def\EPJC(#1,#2,#3){Eur.\ Phys.\ J.\ C\ \issue(#1,#2,#3)}
\def\EPJD(#1,#2,#3){Eur.\ Phys.\ J. Direct\ C\ \issue(#1,#2,#3)}
\def\IEEETNS(#1,#2,#3){IEEE Trans.\ Nucl.\ Sci.\ \issue(#1,#2,#3)}
\def\IJMP(#1,#2,#3){Int.\ J.\ Mod.\ Phys. \issue(#1,#2,#3)}
\def\JHEP(#1,#2,#3){J.\ High Energy Physics \issue(#1,#2,#3)}
\def\JPG(#1,#2,#3){J.\ Phys.\ G \issue(#1,#2,#3)}
\def\MPL(#1,#2,#3){Mod.\ Phys.\ Lett.\ \issue(#1,#2,#3)}
\def\NP(#1,#2,#3){Nucl.\ Phys.\ \issue(#1,#2,#3)}
\def\NIM(#1,#2,#3){Nucl.\ Instrum.\ Meth.\ \issue(#1,#2,#3)}
\def\PL(#1,#2,#3){Phys.\ Lett.\ \issue(#1,#2,#3)}
\def\PRD(#1,#2,#3){Phys.\ Rev.\ D \issue(#1,#2,#3)}
\def\PRL(#1,#2,#3){Phys.\ Rev.\ Lett.\ \issue(#1,#2,#3)}
\def\SJNP(#1,#2,#3){Sov.\ J. Nucl.\ Phys.\ \issue(#1,#2,#3)}
\def\ZPC(#1,#2,#3){Zeit.\ Phys.\ C \issue(#1,#2,#3)}

\usepackage[dvips]{color}
\usepackage[normalem]{ulem}
\def\hlak#1{\textcolor{red}{\large\textsf{#1}}}
\definecolor{darkgreen}{cmyk}{1,0,1,0.4}
\definecolor{pink}{cmyk}{0.4,1,0.3,0}

\begin{document}

\renewcommand*{\thefootnote}{\fnsymbol{footnote}}

\begin{center}
 {\Large\bf{
On the scalar potential of two-Higgs doublet models}}

\vspace{5mm}

{\large Indrani Chakraborty} \footnote{indrani300888@gmail.com} 
and
{\large Anirban Kundu} \footnote{anirban.kundu.cu@gmail.com}\\
{\em{Department of Physics, University of Calcutta, \\
92 Acharya Prafulla Chandra Road, Kolkata 700009, India
}}
\end{center}
\begin{abstract}

We perform a detailed analysis of the Two-Higgs Doublet Model (2HDM) potential. At the tree-level, the 
potential may accommodate more than one minima, one of them being the electroweak (EW) minimum 
where the universe lives. The parameter space allowed after the data from the Large Hadron 
Collider (LHC) came in almost excludes those cases where the EW vacuum is shallower than the second
minimum. We extend the analysis by including terms in the 2HDM potential that break the $Z_2$ symmetry 
of the potential by dimension-4 operators and show that the conclusions remain unchanged. Furthermore, 
a one-loop analysis of the potential is performed for both cases, namely, where the $Z_2$ symmetry of the 
potential is broken by dimension-2 or dimension-4 operators. For quantitative analysis, we show our results 
for the Type-II 2HDM, qualitative results remaining the same for other 2HDMs. We find that the nature of the 
vacua from the tree-level analysis does not change; the EW vacuum still remains deeper. 

\end{abstract}


\date{\today}



\setcounter{footnote}{0}
\renewcommand*{\thefootnote}{\arabic{footnote}}


\section{Introduction}

Two-Higgs Doublet Models (2HDM) \cite{Branco:2011iw,Bhattacharyya:2015nca} 
are one of the most popular extensions of the Standard Model (SM), 
even without invoking supersymmetry, for which more than one scalar doublet is a necessary condition. 
In a 2HDM with two doublets $\Phi_1$ and $\Phi_2$, 
there are five physical scalars: the two $CP$-even neutrals $h$ and $H$, the $CP$-odd neutral
$A$ \footnote{
The neutral scalars do not have any definite $CP$ property if the scalar 
potential violates $CP$. For 2HDM with spontaneous $CP$ violation, see, {\em e.g.}, Ref.\
\cite{Branco:1985aq}.}, and two charged scalars $H^\pm$. While a generic 2HDM may contain 
flavor-changing neutral current (FCNC) interactions (see, {\em e.g.}, Ref.\ \cite{bgl,bgl2,Bhattacharyya:2014nja} 
for such type of 2HDMs), one usually invokes some discrete symmetry to banish such FCNC at the tree-level, based 
on the Glashow-Weinberg-Paschos (GWP) \cite{GWP} theorem that there is no tree-level FCNC 
if all right-handed fermions of a given electric charge couple to only one of the doublets.
It turns out that there are four 2HDMs that satisfy the GWP criterion when a discrete $Z_2$ symmetry 
is applied on the Lagrangian. They are:
\begin{enumerate}
 \item  Type I, for which all fermions couple with $\Phi_2$ and none with $\Phi_1$;
 \item  Type II, for which up-type quarks couple to $\Phi_2$, down-type quarks and charged leptons couple to $\Phi_1$;
 \item  Type Y (sometimes called Type III or Flipped), for which up-type quarks and
charged leptons couple to $\Phi_2$ and down-type quarks couple to $\Phi_1$;
 \item Type X (sometimes called Type IV or Lepton-specific), for which all charged leptons couple to $\Phi_1$ and 
all quarks couple to $\Phi_2$.
\end{enumerate}
Among them, Type II 2HDM has been most widely investigated 
\cite{Cheon:2012rh,Coleppa:2013dya,Eberhardt:2013uba,Eberhardt:2014kaa,Chakrabarty:2014aya} 
because the scalar sector of minimal 
supersymmetry is a Type II 2HDM. 

The tightest constraint on any 2HDM comes from the fact that the observed scalar resonance at the Large Hadron 
Collider (LHC) can be identified with the SM Higgs boson with $m_{\rm Higgs}=125.09\pm 0.24$ GeV 
\cite{Aad:2015zhl}.
Thus, the Yukawa and gauge couplings of the 2HDM must be so aligned as to make the 
lighter $CP$-even mass eigenstate $h$ to almost coincide with the SM Higgs boson. This is known as the 
{\em alignment limit}, and while the allowed parameter space may vary from one 2HDM to the other, the 
qualitative results are quite similar \cite{alignment,1507.00933,Das:2015mwa}. 
There are other constraints, like the oblique parameters 
or the lower limit on the $H^\pm$ mass coming 
from $b\to s\gamma$ decay rate \cite{Mahmoudi:2009zx,Deschamps:2009rh}, but 
such constraints are in general not equally valid for all 2HDMs \footnote{
For example, 
the $b\to s\gamma$ constraint, $m_{H^\pm} > 316$ GeV, can be evaded in Type I and Lepton-specific 2HDM.}. 
The theoretical constraints include, 
just like any other extensions of the SM, the vacuum stability, validity of the perturbative nature of the couplings, 
and constraints coming from the requirement of unitarity of scattering amplitudes \cite{Gorczyca:2011he}.

In this paper, we will focus upon the scalar potential of the 2HDMs. This is much more complicated, even at the 
tree-level, compared to the SM scalar potential, if both the $CP$-even neutral scalars are allowed to have nonzero 
vacuum expectation values (VEV). As has been shown in Refs.\ 
\cite{Barroso:2013awa,Barroso:2013kqa,Ferreira:2004yd,Ivanov:2006yq,Ivanov:2007de},
the 2HDM scalar potential, even with the softly broken $Z_2$-symmetry, may allow more than one minima. 
A similar analysis was very recently done for $Z_2$-breaking models \cite{Ivanov:2015nea}. 
In Refs.\  \cite{Barroso:2013awa,Ferreira:2004yd,Ivanov:2006yq,Ivanov:2007de}, 
the authors considered the scalar potential of a $Z_2$-conserving 2HDM at the tree-level, and showed that 
it can allow multiple normal non-equivalent stationary points (at most two of them can be minima).
However, a charge-violating or $CP$-violating minimum cannot coexist with a
normal minimum, {\em i.e.}, where the $CP$-even neutral fields get the VEV. This can be put in a more succinct way: 
minima with different natures cannot co-exist in 2HDM \cite{Barroso:2013awa,Ivanov:2007de}.
Also, the data from LHC all but rules 
out those points where the second minimum is deeper than the electroweak (EW) minimum, {\em i.e.}, the minimum
where the universe lives. 

In this paper, we consider both $Z_2$-conserving and $Z_2$-violating 2HDM scalar potentials for our analysis. By $Z_2$-breaking 
2HDM, we mean those with dimension-4 operators violating $Z_2$, like $(\Phi_1^\dag\Phi_1)(\Phi_1^\dag\Phi_2)$ or  $(\Phi_2^\dag\Phi_2)(\Phi_1^\dag\Phi_2)$. 
There can be soft $Z_2$-violating terms like $\Phi_1^\dag \Phi_2$ or $\Phi_2^\dag \Phi_1$ in the potential; if there are no dimension-4 
$Z_2$-breaking terms, we call those models $Z_2$-conserving, although, strictly speaking, they are not.
We show that the conclusions drawn about the nature of the 
minima of the potential in the context of $Z_2$-conserving 2HDMs remain valid for $Z_2$-violating 2HDMs too. 

To check the robustness of the tree-level results, 
we further perform a one-loop analysis of the 2HDM potential, and choose only those models that show a double minima. 
A nice review of the one-loop corrections in the context of $Z_2$-symmetric 2HDM and scale invariant 2HDM can be found in
Refs.\ \cite{Sher:1988mj,Sher:1996ib,Lee:2012jn}. 
It is a common knowledge that the one-loop corrections \cite{Coleman:1973jx} can be significant only in the flat direction of the potential. 
For the SM, this is easy to obtain \cite{Quiros:1999jp}; so is for the Inert Doublet models \footnote{ 
For the Inert Doublet model (which is important from the cosmological implication of providing a cold dark matter candidate),
one of the VEVs is zero, and so it is easier to treat analytically. 
We, however, will not go into any detailed study of such cosmological implications of the 2HDMs in this paper.
Another such implication is the successful first-order electroweak phase transition and electroweak baryogenesis,
which is discussed in Refs.\ \cite{baryogenesis}.}
\cite{Khan:2015ipa,Swiezewska:2015paa} where one of the VEVs is zero. For a 
generic 2HDM, this is a cumbersome task, but can be done, in principle, following the prescriptions of Gildener and 
Weinberg \cite{Gildener:1976ih}, and the ray in the potential space along which the tree-level potential is zero can be found.  
However, If $v_2 \gg v_1$, where $v_1$ and $v_2$ are the VEVs of the $CP$-even neutral components of $\Phi_1$ and 
$\Phi_2$ respectively (so that $\tan\beta \equiv v_2/v_1 \gg 1$), the potential along the $\Phi_1$ direction is {\em
almost} flat, so it is instructive to show the variations of the potential perpendicular to this direction, {\em i.e.},
along $\Phi_2$. Another important point is
the setting of the regularization scale $\mu$ for the one-loop corrections. Variation of $\mu$ is equivalent to the variations
of the tree-level quartic couplings $\lambda_i$, as can be seen from a renormalization group argument. As the nature 
of the potential can best be described by these quartic couplings, we would like to forward a prescription of choosing 
$\mu$ for the 2HDM: choose it so that the position of the EW minimum remains unaltered. As we will see, this keeps the
position of the second minimum too almost unaltered. Of course, the depths of the potential at the two minima will change. 
With such a prescription for choosing $\mu$, the conclusions about the stability of the EW minimum that were drawn from 
a tree-level analysis remain unaltered.

The paper is arranged as follows. 
In section II we briefly review 2HDMs,  with softly broken $Z_2$ symmetry and 
without $Z_2$ symmetry, and list the constraints and the minimization conditions on the potential. 
Section III introduces the one-loop corrected effective potential and modified minimization conditions for Type II 2HDM 
with and without $Z_2$ symmetry. Our results, for both tree-level and one-loop analysis, are shown in Section IV. 
Section V summarizes the paper. Some calculation details and relevant expressions are relegated to the 
Appendix.

\section{Brief review of 2HDM}

\subsection{2HDM with softly broken $Z_2$ symmetry}

To start with, let us focus on the most canonical 2HDMs, where the $Z_2$ symmetry is broken 
only softly by a dimension-2 operator, and all the couplings are real. The notations used here essentially 
follow those in Ref.\ \cite{Branco:2011iw}.  Later on, we will introduce both dimension-4 $Z_2$-breaking operators 
as well as complex parameters in the scalar potential.

Let us denote the two doublets, both with hypercharge $Y=+1$, by $\Phi_1$ and $\Phi_2$, which can be 
written more explicitly as 
\be
\Phi_a= \begin{pmatrix} \chi_a^+ \cr \frac{1}{\sqrt{2}}\left(\phi_a + i\eta_a\right) \end{pmatrix}\,, 
\ \ \ a = 1,2\,.
\ee
We further assume the VEVs to be aligned towards the direction of the 
CP-even neutral field, so that $\bra \phi_1\ket = v_1$, $\bra\phi_2\ket = v_2$, and we conventionally denote 
$\tan\beta = v_2/v_1$. 

Invoking a $Z_2$ symmetry $\Phi_1\to\Phi_1$, $\Phi_2\to -\Phi_2$, so that there is no tree-level 
flavor-changing neutral current (FCNC), one may write
\bea
 V \left(\Phi_1 , \Phi_2\right)&=&  m_{11}^2 \Phi_1^\dag\Phi_1 + m_{22}^2 \Phi_2^\dag\Phi_2 - m_{12}^2 \left(\Phi_1^\dag\Phi_2 
  + \Phi_2^\dag\Phi_1\right)\nonumber\\
  && + \frac12\lambda_1\left(\Phi_1^\dag\Phi_1\right)^2+\frac12\lambda_2\left(\Phi_2^\dag\Phi_2\right)^2
     + \lambda_3 \left(\Phi_1^\dag\Phi_1\right)\left(\Phi_2^\dag\Phi_2\right) \nonumber\\
&&   + \lambda_4 \left(\Phi_1^\dag\Phi_2\right)\left(\Phi_2^\dag\Phi_1\right)+
\frac12\lambda_5\left[\left(\Phi_1^\dag\Phi_2\right)^2 + \left(\Phi_2^\dag\Phi_1\right)^2\right]\,.
\label{Vtreecon}
\eea
Here $m_{11}^2$, $m_{22}^2$, $m_{12}^2$, $\lambda_1$, $\lambda_2$, 
$\lambda_3$, $\lambda_4$ and $\lambda_5$ are all real and $m_{12}^2$ softly breaks the $Z_2$ symmetry. 
As mentioned before, to differentiate 
from the potential that breaks the $Z_2$ symmetry with dimension-4 operators, such models will henceforth be called the 
$Z_2$-conserving or $Z_2$-symmetric models, even though it is broken softly by $m_{12}^2$.

The two CP-even neutral states $\phi_1$ and $\phi_2$ are in general not mass eigenstates. 
The corresponding mass matrix can be diagonalized through 
a rotation by an angle $\alpha$, and the mass eigenstates are
\be
h = \phi_2\cos\alpha - \phi_1\sin\alpha\,,\ \ 
H = \phi_2\sin\alpha + \phi_1\cos\alpha\,,
\ee
where $h (H)$ is the lighter (heavier) eigenstate. 

Note that if $|\alpha -\beta|$ is an odd (even) multiple of $\pi/2$, 
$h(H)$ becomes identical with the SM Higgs boson, with a VEV of $v = \sqrt{v_1^2+v_2^2}\approx 246$ GeV.
For example, the $hVV^\ast$ 
($HVV^\ast$)
coupling is just the SM coupling times $\sin(\alpha-\beta)$ ($\cos(\alpha-\beta)$), where $V$ 
is any weak gauge boson, $W$ or $Z$. 
The limit where $h$ behaves as the SM Higgs boson is known as the 
alignment limit. The LHC data strongly favours the alignment limit and this sets a nontrivial constraint on 
the parameter space. The allowed parameter space, of course, depends on what type of 2HDM is chosen.
We refer the reader to Ref.\ \cite{alignment,1507.00933} for a study of the alignment limit in Type I and Type II 2HDMs,
and to Ref.\ \cite{Bhattacharyya:2014nja} for a typical example of constraints coming from 2HDMs with tree-level FCNC.

The most generic Yukawa interactions for these four models can be written as \cite{Branco:2011iw},
\be
\mathcal{L}_Y = -\sum_{j=1}^2 \left[ Y_j^d \bar{Q}_L d_R \Phi_j + Y_j^u \bar{Q}_L u_R \tilde{\Phi}_j
+ Y_j^e \bar{L}_L l_R \Phi_j + {\rm h.c.}\right] \,,
\ee
where $\tilde{\Phi}_j = i{\tau_2}\Phi_j^*$, $Q_L$, $L_L$, $d_R$, $u_R$ and $l_R$ are 
generic doublet quarks, doublet leptons, singlet down-type and singlet up-type quarks, and singlet 
charged leptons respectively. $Y_j^d$, $Y_j^u$, $Y_j^e$ are $3\times 3$ complex matrices, 
containing Yukawa couplings for the down, up, and leptonic sectors respectively. In our analysis 
we will consider only top, bottom, and $\tau$ Yukawa couplings to be nonzero. 

The masses of the charged Higgs, $H^\pm$, and the pseudoscalar, $A$, can be written as
\bea
m_{H^\pm}^{2} &=& \frac{1}{v_1 v_2}m_{12}^2 v^2 - \frac12 \left(\lambda_4 + \lambda_5\right)v^2\,,\nonumber\\
m_A^2 &=& \frac{1}{v_1 v_2}m_{12}^2 v^2 - \lambda_5 v^2\,.
\label{masssq-z2}
\eea

The couplings can be expressed in terms of masses of the physical states and the mixing angles 
$\alpha$ and $\beta$ as \cite{Barroso:2013awa} 
\bea
\lambda_1 &=& \frac{1}{v^2 c_{\beta}^2}\left(c_{\alpha}^2 m_H^2 + s_{\alpha}^2 m_h^2 - \frac{m_{12}^2 s_{\beta}} {c_{\beta}}\right) \,,\nonumber\\
\lambda_2 &=& \frac{1}{v^2 s_{\beta}^2}\left(s_{\alpha}^2 m_H^2 + c_{\alpha}^2 m_h^2 - \frac{m_{12}^2 c_{\beta}} {s_{\beta}}\right) \,,\nonumber\\
\lambda_3 &=& \frac{2 m_{H^\pm}^2}{ v^2} + \frac{s_{2\alpha}}{v^2 s_{2\beta}} \left(m_H^2 - m_h^2\right) - \frac{m_{12}^2}{v^2 s_\beta c_\beta}\,, \nonumber\\
\lambda_4 &=& \frac{1}{v^2} \left(m_A^2 - 2m_{H^\pm}^2\right) + \frac{m_{12}^2}{v^2 s_\beta c_\beta}\,, \nonumber\\
\lambda_5 &=& \frac{m_{12}^2}{v^2 s_\beta c_\beta} - \frac{m_A^2}{v^2}\,.
\eea
where $m_h (m_H)$ is the mass of $h (H)$, and 
$c_\theta$ and $s_\theta$ are generic shorthand notations for $\cos\theta$ and $\sin\theta$ 
respectively.

The requirement that the scalar potential always remains bounded from below leads to the 
following stability conditions for 2HDMs with $Z_2$ symmetry \cite{Branco:2011iw},
\be
\lambda_1\,,\lambda_2 \ge 0\,,\ \ 
\lambda_3\geq -{\sqrt{\lambda_1\lambda_2}}\,,\ \  
\lambda_3 + \lambda_4 - \vert\lambda_5\vert \geq  -{\sqrt{\lambda_1\lambda_2}}\,.
\label{stab-z2}
\ee
Note that $\lambda_3$, $\lambda_4$, and $\lambda_5$ can in principle be negative; however, if the $Z_2$ symmetry is 
exact, this may lead to tachyonic masses for $H^\pm$ and $A$.


\subsection{2HDM with hard ${Z_2}$ breaking terms}

The most general renormalizable scalar potential of 2HDM can be written as \cite{Branco:2011iw}
\bea
 V \left(\Phi_1 , \Phi_2\right) &=&  m_{11}^2 \Phi_1^\dag\Phi_1 + m_{22}^2 \Phi_2^\dag\Phi_2 - 
 \left[m_{12}^2 \left(\Phi_1^\dag\Phi_2 \right)+h.c.\right]\nonumber\\
  && + \frac12\lambda_1\left(\Phi_1^\dag\Phi_1\right)^2+\frac12\lambda_2\left(\Phi_2^\dag\Phi_2\right)^2
     + \lambda_3 \left(\Phi_1^\dag\Phi_1\right)\left(\Phi_2^\dag\Phi_2\right) + \lambda_4 
     \left(\Phi_1^\dag\Phi_2\right)\left(\Phi_2^\dag\Phi_1\right)\nonumber\\
  &&   +\left[\frac12\lambda_5\left(\Phi_1^\dag\Phi_2\right)^2+
\lambda_6\left(\Phi_1^\dag\Phi_1\right)\left(\Phi_1^\dag\Phi_2\right) 
 + \lambda_7\left(\Phi_2^\dag\Phi_2\right)\left(\Phi_1^\dag\Phi_2\right) + h.c.\right]\,,
 \label{Vtreebreak}
\eea
where the model parameters $m_{11}^2$, $m_{22}^2$, $\lambda_1$, $\lambda_2$, $\lambda_3$, 
$\lambda_4$ are real and $m_{12}^2$, $\lambda_5$, $\lambda_6$ and $\lambda_7$
can in principle be complex, and ``h.c.'' stands for hermitian conjugation. 

We will follow the same convention as the previous subsection and consider two different cases, namely,
$m_{12}^2$, $\lambda_5$, $\lambda_6$ and $\lambda_7$ are (i) real, and (ii) complex.

\subsection{2HDMs without $Z_2$ symmetry with all real parameters}

For this case, the masses of $H^\pm$ and $A$ can be written in an analogous way of Eq.\ (\ref{masssq-z2}), 
using the minimization conditions of the potential:
\bea
m_{H^\pm}^2 &=& \frac{1}{v_1 v_2}m_{12}^2 v^2 - 
\frac12 \left(\lambda_4 + \lambda_5\right)v^2 -\frac{v_1}{2 v_2} \lambda_6 v^2 -\frac{v_2}{2 v_1} \lambda_7 v^2\,,
\nonumber\\
m_A^2 &=& \frac{1}{v_1 v_2}m_{12}^2 v^2 - \lambda_5 v^2 -\frac{v_1}{2 v_2} \lambda_6 v^2 -\frac{v_2}{2 v_1} \lambda_7 v^2\,.
\label{massq-noz2-real}
\eea
The requirement that the scalar potential always remains bounded from below leads to the 
same set of equations as in (\ref{stab-z2}), with an extra condition:
\bea
&&\lambda_1\,,\lambda_2 \ge 0\,,\ \ 
\lambda_3\geq -{\sqrt{\lambda_1\lambda_2}}\,,\ \  
\lambda_3 + \lambda_4 - \vert\lambda_5\vert \geq  -{\sqrt{\lambda_1\lambda_2}}\,,\nonumber\\
&&2\vert \lambda_6 + \lambda_7\vert \le \frac12\left(\lambda_1 + \lambda_2\right) + \lambda_3 + \lambda_4 + \lambda_5\,.
\label{stab-noz2-1}
\eea
The mixing angle between the CP-even neutral states is given by $\alpha=\frac12\arctan(A/B)$, where
\bea
A &=& m_{12}^2 - \left(\lambda_3 + \lambda_4 + \lambda_5\right) v_1 v_2 - \frac32 \lambda_6 v_1^2 - \frac32 \lambda_7 v_2^2\nonumber\\
B &=& - \frac12 \lambda_1 v_1^2 + \frac12 \lambda_2 v_2^2 + \frac12 m_{12}^2 \left(v_1/v_2 - v_2/v_1\right)
+ \frac34\left(\lambda_7-\lambda_6\right) v_1 v_2 + \frac{1}{4v_1} \lambda_7 v_2^3 -  \frac{1}{4v_2}\lambda_6 v_1^3\,.
\eea

\subsubsection{Minima of the potential}

The scalar potential of the 2HDM shows a much more complicated structure 
than that of the SM. The potential can have multiple non-equivalent normal stationary points 
(only two of them can be minima) 
\cite{Barroso:2013awa,Ferreira:2004yd,Ivanov:2006yq,Ivanov:2007de}, depending on the parameters. These minima can be 
all normal (where only $\phi_1$ and $\phi_2$ get nonzero VEV), charge-breaking (where at least one of the charged 
fields $\chi^+_a$ gets a nonzero VEV) or CP violating (where the two VEVs have a nontrivial phase between them). It has been 
shown in Refs.\ \cite{Barroso:2013awa,Barroso:2013kqa,Ferreira:2004yd,Ivanov:2007de}, 
that for the canonical $Z_2$-symmetric 2HDM, (i) existence of a normal minimum 
rules out a charge breaking or CP violating minimum, those extrema can only be saddle points at best; (ii) there 
can be more than one normal minima, only one of which corresponds to the Standard model (SM), which we call the EW minimum. 
If the EW minimum is the global one, 
we are in a stable situation; if it is not, the universe may tunnel down to the deeper minimum if the tunneling 
time is less than or of the order of the lifetime of the universe. It has also been shown that the LHC 
data effectively rules out the parameter space where the EW minimum is shallower. We would like to extend this result to
$Z_2$-breaking 2HDM with both real and complex parameters, and also investigate the nature of the potential 
when one-loop corrections are taken into account.

In a normal minimum, we may write
\be
\bra\Phi_1\ket_N = \frac{1}{\sqrt{2}} \begin{pmatrix} 0 \cr v_1 \end{pmatrix}\,, 
\bra\Phi_2\ket_N = \frac{1}{\sqrt{2}} \begin{pmatrix} 0 \cr v_2 \end{pmatrix}\, .
\ee
\hlak 
In certain situations, several non-equivalent normal extrema are allowed by the minimization 
conditions of the potentials and at most two of them can be local minima.
Thus, there can be a second minimum with VEVs $(v'_1,v'_2)$, with ${v'} = \sqrt{{v'}_1^2 + {v'}_2^2} 
\not= 246$ GeV, where exactly same symmetries are broken. 
Obviously, even with the same parameters of the potential, the masses of 
all SM particles are going to change from the EW vacuum configuration. 

At the EW vacuum, the potential, as follows from Eq.(\ref {Vtreebreak}), can be written as
\bea
V_0 &=& \frac12 m_{11}^2 v_1^2 + \frac12 m_{22}^2 v_2^2 - m_{12}^2 v_1v_2 + \frac18 \lambda_1 v_1^4 + \frac18 \lambda_2 v_2^4
+ \frac14 \left(\lambda_3 + \lambda_4 + \lambda_5\right)v_1^2 v_2^2\nonumber\\
&&+ \frac 12 \lambda_6 v_1^3 v_2 + \frac12 \lambda_7 v_2^3 v_1\,,
\label{sm-vac-pot}
\eea
and at the other minimum, we get the corresponding $V'_0$ by
replacing ($v_1$,$v_2$) in Eq.\ (\ref{sm-vac-pot}) by $(v'_1,v'_2)$.

If $V_0 < (>)  V'_0$, the EW (second) vacuum is stable and is the global minimum. 

The minimization conditions of the potential are easy to obtain:
\bea
f_1(v_1,v_2) &\equiv& m_{11}^2v_1 - m_{12}^2v_2 +\frac12 \lambda_1v_1^3 + 
\frac12 \left(\lambda_3 + \lambda_4 + \lambda_5\right)v_1v_2^2 + \frac32 \lambda_6 v_1^2 v_2
+ \frac 12 \lambda_7 v_2^3 = 0\,, \nonumber\\
f_2(v_1,v_2) &\equiv& m_{22}^2v_2 - m_{12}^2v_1 +\frac12 \lambda_2v_2^3 + 
\frac12 \left(\lambda_3 + \lambda_4 + \lambda_5\right)v_2v_1^2  + 
\frac 12 \lambda_6 v_1^3 + \frac32 \lambda_7 v_2^2 v_1 = 0\,.
\label{potmin-1}
\eea
with
\bea
\left[m_{11}^2 + \frac{3}{2} \lambda_1 v_1^2 + 
\frac12 \left(\lambda_3 + \lambda_4 + \lambda_5\right)v_2^2 + 3\lambda_6 v_1 v_2\right] &>& 0\,, 
\nonumber \\
\left[m_{22}^2 + \frac{3}{2} \lambda_2 v_2^2 + 
\frac12 \left(\lambda_3 + \lambda_4 + \lambda_5\right)v_1^2 + 3 \lambda_7 v_1 v_2\right] &>& 0 \,.
\eea

\subsubsection{Charge-breaking minimum}

If the symmetry breaks in such a way that one of the charged fields, say $\chi_2^+$, gets a non-zero VEV $\gamma$ too, 
such that
\be
\bra\Phi_1\ket_{CB} = \frac{1}{\sqrt{2}} \begin{pmatrix} 0 \cr v_1 \end{pmatrix}\,, 
\bra\Phi_2\ket_{CB} = \frac{1}{\sqrt{2}} \begin{pmatrix} \gamma \cr v_2 \end{pmatrix}\,,
\ee
the minimization conditions of the potential are 
\bea
&& m_{11}^2v_1 - m_{12}^2v_2 +\frac12 \lambda_1v_1^3 + \frac12 \lambda_3 v_1 \left(v_2^2 + \gamma^2\right) 
+\frac12 \left( \lambda_4 + \lambda_5\right)v_1v_2^2
+ \frac32 \lambda_6 v_1^2 v_2 + \frac12 \lambda_7\left(\gamma^2 + v_2^2\right)v_2 = 0\,, \nonumber\\
&&m_{22}^2v_2 - m_{12}^2v_1 +\frac12 \lambda_2 \left(v_2^2+\gamma^2\right)v_2 + 
\frac12 \left(\lambda_3 + \lambda_4 + \lambda_5\right)v_2v_1^2 
+ \frac12 \lambda_6 v_1^3 + \frac12 \lambda_7 \left(\gamma^2 + 3 v_2^2\right)v_1 = 0\,, \nonumber\\
&&m_{22}^2 + \frac12 \lambda_2 \left(\gamma^2 + v_2^2\right) + \frac12 \lambda_3 v_1^2 + \lambda_7 v_1 v_2 = 0 \,,
\label{potmin-cb}
\eea
and 
\bea
m_{11}^2 +  \frac32 \lambda_1 v_1^2 + 
\frac12 \lambda_3\left(\gamma^2+v_2^2\right) + 
\frac12 \left(\lambda_4 + \lambda_5\right)v_2^2+3 \lambda_6 v_1 v_2 &>& 0\,, \nonumber\\
m_{22}^2 + \frac12 \lambda_2 \left(3 v_2^2 + \gamma^2\right) + 
\frac12 \left(\lambda_3 + \lambda_4 + \lambda_5\right)v_1^2 + 3 \lambda_7 v_1 v_2 &>& 0\,, \nonumber\\
m_{22}^2 + \frac12 \lambda_2 \left(3 \gamma^2 + v_2^2\right) + \frac14 \lambda_3 v_1^2 + 
\lambda_7 v_1 v_2 &>& 0\,.
\label{potmin-cb2}
\eea

\subsubsection{CP violating minimum}
At the CP violating extremum, one of the CP-odd neutral fields acquire a nonzero 
VEV $\delta$. In other words, the VEVs develop a relative phase.  
Thus,
\be
\Phi_1 = \begin{pmatrix} {\chi_1}^+ \cr \frac{1}{\sqrt{2}}\left(\left(\phi_1 + 
v_1\right) + i\left(\eta_1 + \delta\right)\right) \end{pmatrix}\,,\ \
\Phi_2= \begin{pmatrix} {\chi_2}^+ \cr \frac{1}{\sqrt{2}}\left(\left(\phi_2 + 
v_2\right) + i\eta_2\right) \end{pmatrix}\,.
\ee

 We have similar minimization conditions for the potential: 
\bea
&& m_{11}^2v_1 - m_{12}^2v_2 +\frac12 \lambda_1\left(v_1^2 + \delta^2\right) v_1  +\frac12 \left(\lambda_3 + \lambda_4 + \lambda_5\right)v_1v_2^2
+ \frac12 \lambda_6\left(3v_1^2 + \delta^2\right)v_2 + \frac12 \lambda_7 v_2^3= 0\,, \nonumber\\
&& m_{22}^2v_2 - m_{12}^2v_1 +\frac12 \lambda_2 v_2^3 + \frac12 \left(\lambda_3 + \lambda_4\right) v_2\left(v_1^2 + \delta^2\right) 
+ \frac12 \lambda_5 \left(v_1^2 -\delta^2\right) v_2 
 + \frac12 \lambda_6 \left(v_1^2 +\delta^2\right) v_1 \nonumber\\
 && \ \ \ \ \ \ \ \ \ \ \ \ + \frac32 \lambda_7 v_1 v_2^2  = 0\,, \nonumber\\
&& m_{11}^2 \delta + \frac12 \lambda_1 \left(\delta^2 + v_1^2\right)\delta + \frac12 \left(\lambda_3 + \lambda_4\right)v_2^2 \delta - \frac12 \lambda_5 v_2^2 \delta
+ \lambda_6 v_1 v_2 \delta = 0 \,,
\eea 
and
\bea
m_{11}^2 + \frac12 \lambda_1\left(3 v_1^2 + \delta^2\right) +  \frac12 \left(\lambda_3 + \lambda_4 + \lambda_5\right)v_2^2
+ 3 \lambda_6 v_1 v_2 &>& 0\,, \nonumber\\
m_{22}^2 + \frac 32 \lambda_2 v_2^2 + \frac 12 \left(\lambda_3 + \lambda_4\right) \left(v_1^2 +\delta^2\right) 
+ \frac12 \lambda_5 \left(v_1^2 -\delta^2\right) + 3 \lambda_7 v_1 v_2 &>& 0\,, \nonumber\\
m_{11}^2 + \frac12 \lambda_1 \left(3v_1^2 + \delta^2\right) + 
\frac12 \left(\lambda_3 + \lambda_4 - \lambda_5\right)v_2^2 + \lambda_6 v_1 v_2 &>& 0 \,.
\eea

\subsection{2HDM without $Z_2$ symmetry with complex parameters}

Next we consider the 2HDMs with $m_{12}^2$, $\lambda_5$, $\lambda_6$ and $\lambda_7$ complex, and use 

\be
m_{12}^2 = m_{12}^{2R} + i m_{12}^{2I}\,,\ \ 
\lambda_5 = \lambda_{51} + i \lambda_{52}\,,\ \ 
\lambda_6= \lambda_{61} + i \lambda_{62}\,, \ \ 
\lambda_7 = \lambda_{71} + i \lambda_{72} \,.
\ee
Most of the relevant expressions are identical with the real parameter case discussed before, with 
the $(m_{12}^{2R},\lambda_{51},\lambda_{61},\lambda_{71})$ replacing $(m_{12}^2,\lambda_5,\lambda_6,\lambda_7)$ 
respectively. While this applies to the expressions of physical mass eigenstates, there is an extra condition 
so that the charged Goldstone boson is still massless:
\be
-m_{12}^{2I} + \frac12 \lambda_{52} v_1 v_2 + \frac12 \lambda_{62} v_1^2 + \frac12 \lambda_{72} v_2^2 = 0\,.
\label{zero-gold}
\ee
This shows that only three out of the four phases are independent.

The conditions for the stability of the potential are 
\bea
&&\lambda_1\,,\lambda_2 \ge 0\,,\ \ 
\lambda_3\geq -{\sqrt{\lambda_1\lambda_2}}\,,\ \ 
\lambda_3 + \lambda_4 - {\sqrt{\lambda_{51}^2 + \lambda_{52}^2}} \geq  -{\sqrt{\lambda_1\lambda_2}}\,,\nonumber\\
&&2\vert \lambda_{61} + \lambda_{71}\vert \le \frac12\left(\lambda_1 + \lambda_2\right) + \lambda_3 + \lambda_4 + \lambda_{51}
\,,\ \ 
2\vert \lambda_{62} + \lambda_{72}\vert \le \frac12\left(\lambda_1 + \lambda_2\right) + \lambda_3 + \lambda_4 - \lambda_{51}\,.
\eea

Similarly, the minimization conditions for the normal and charge-breaking minima are obtained from the corresponding 
expressions for the real parameter case by the substitutions mentioned before. For the charge-breaking minima, 
the imaginary parts of the potential parameters play a role. 
 The minimization conditions are
\bea
m_{11}^2v_1 - m_{12}^{2R} v_2 +\frac12 \lambda_1\left(v_1^2 + \delta^2\right) v_1 +\frac12 \left(\lambda_3 + \lambda_4 + \lambda_{51}\right)v_1v_2^2
+\frac12 \lambda_{52}v_2^2 \delta  && \nonumber\\
+ \frac12 \lambda_{61}\left(3v_1^2+ \delta^2\right)v_2 + \lambda_{62} v_1 v_2 \delta +  \frac12 \lambda_{71} v_2^3 &=& 0\,, \nonumber\\
m_{22}^2v_2 - m_{12}^{2R}v_1 - m_{12}^{2I} \delta + \frac12 \lambda_2 v_2^3 + \frac12 \left(\lambda_3 + \lambda_4\right) v_2\left(v_1^2 + \delta^2\right) + \frac12 \lambda_{51} \left(v_1^2 -\delta^2\right) v_2 && \nonumber\\
+ \lambda_{52} v_1 v_2 \delta + \frac12 \lambda_{61} \left(v_1^2 +\delta^2\right) v_1 + \frac12 \lambda_{62} \left(v_1^2 +\delta^2\right) \delta + \frac32 \lambda_{71} v_1 v_2^2
 + \frac32 \lambda_{72}v_2^2 \delta &=& 0\,, \nonumber\\
m_{11}^2 \delta -m_{12}^{2I} v_2 + \frac12 \lambda_1 \left(\delta^2 + v_1^2\right)\delta 
+ \frac12 \left(\lambda_3 + \lambda_4\right)v_2^2 \delta - \frac12 \lambda_{51} v_2^2 \delta
+\frac12\lambda_{52}v_1 v_2^2 && \nonumber\\
+ \lambda_{61} v_1 v_2 \delta + \frac12 \lambda_{62}\left(3\delta^2 + v_1^2\right) v_2 
+ \frac12 \lambda_{72}v_2^3 &=& 0 \,,
\eea
and 
\bea
m_{11}^2 + \frac12 \lambda_1\left(3 v_1^2 + \delta^2\right) +  \frac12 \left(\lambda_3 + \lambda_4 + \lambda_{51}\right)v_2^2
+ 3 \lambda_{61} v_1 v_2 + \lambda_{62} v_2 \delta  &>& 0\,, \nonumber\\
m_{22}^2 + \frac 32 \lambda_2 v_2^2 + \frac 12 \left(\lambda_3 + \lambda_4\right) \left(v_1^2 +\delta^2\right) 
+ \frac12 \lambda_{51} \left(v_1^2 -\delta^2\right)+ \lambda_{52}v_1 \delta + 3 \lambda_{71} v_1 v_2 + 3 \lambda_{72} v_2 \delta 
&>& 0\,, \nonumber\\
m_{11}^2 + \frac12 \lambda_1 \left(3v_1^2 + \delta^2\right) + \frac12 \left(\lambda_3 + \lambda_4 - \lambda_{51}\right)v_2^2 + \lambda_{61} v_1 v_2 + 3\lambda_{62} v_2 \delta &>& 0 \,.
\eea

The tree-level analysis will be performed using these expressions and the corresponding solutions of the simultaneous 
minimization equations. 

\section{One-loop effective potential}

The form of the one-loop correction to the scalar potential is quite standard, and is given by
\be
m_i \to m_i(\phi_{c1},\phi_{c2})\,,\ \ \ \ 
V_1 = \frac{1}{64\pi^2} \sum_{i=B,F} N_i m_i^4\left( \ln \frac{m_i^2}{\mu^2} - C_i\right)\,,
\label{1loop-generic}
\ee
where the sum runs over all bosonic (the physical scalars, the unphysical Goldstone bosons, $W$ and $Z$) and fermionic 
(for our case, only $t$, $b$ and $\tau$) degrees of freedom. Here $\mu$ is the regularization scale. 
The masses are field-dependent quantities, being functions of the 
background fields $\phi_{c1}$ and $\phi_{c2}$. They can be thought as the positions of the minima in the field space. In the 
limit $\phi_{c1}=v_1$, $\phi_{c2}=v_2$, the field-dependent masses are equal to the physical masses. The constants $N_i$ and $C_i$ 
are given by
\bea
&&N_h = N_H = N_A = N_{G^0} = 1\,,\ \ N_{H^\pm} = N_{G^\pm} = 2\,,\nonumber\\
&&N_W = 6\,,\ \ N_Z = 3\,,\ \ 
N_t = N_b = -12\,,\ \ N_\tau = -4\,,\nonumber\\
&&C_h = C_H = C_A = C_{H^\pm} = C_{G^0} = C_{G^\pm} = C_t = C_b = C_\tau = \frac32\,,\ \ 
C_W = C_Z = \frac56\,.
\eea
The full potential can be written as
\be
V = V(\Phi_1,\Phi_2) + V_1
\label{V_eff}
\ee
where $V(\Phi_1,\Phi_2)$ is the tree-level potential as given in Eq. (\ref{Vtreecon}) or (\ref{Vtreebreak}).

At this point, let us again note that the masses in Eq.\ (\ref{1loop-generic}) are functions of background fields 
$\phi_{c1}$ and $\phi_{c2}$. The positions of the minima are functions of not only $\phi_{c1}$ and $\phi_{c2}$ 
but also the regularization scale $\mu$. To have any idea of the nature of the potential after one-loop correction, one has 
to fix $\mu$ by some prescription. The renormalization group improved one-loop potential clearly shows that 
the variation in $\mu$ is equivalent to the redefinition of the coupling parameters of the theory. 
One popular way is to fix it in such a way that the one-loop corrections are minimum, hoping 
that this will minimize the higher-order corrections too. To be physically more transparent, we try a different approach: we fix $\mu$ 
in such a way that the position of the EW minimum remains (almost) unchanged with respect to the 
tree-level position. This will keep the one-loop corrected field-dependent 
masses to be at the same values of the tree-level masses 
at the EW vacuum. Only the depth of the potential changes by the one-loop 
correction. Thus, our prescription is to tune $\mu$ in such a way that 
$\phi_{c1}=v_1$ and $\phi_{c2}=v_2$ (so that, for example, the Goldstone bosons are still massless \footnote{The
treatment of the Goldstone bosons in one-loop corrected potentials is tricky, and a consistent treatment 
needs resummation of the Goldstone contributions in the effective potential 
\cite{Martin:2014bca}.}). As we will show
later, if one tunes $\mu$ so that the position of the EW minimum is unchanged, it will keep the position of the 
second minimum almost unchanged too. 

As a concrete example, let us now discuss the Type II 2HDM; as the Yukawa couplings enter the picture, one needs to
specify the type of 2HDM under consideration. We have checked that the qualitative features remain unchanged in all other 2HDMs. 

\subsection{Type-II 2HDM without $Z_2$: One-loop corrected potential}

To keep the discussion as much general as possible, let us focus on 
the one-loop correction to the 2HDM (without $Z_2$ symmetry) tree level potential. We can obtain the same
for $Z_2$ symmetric 2HDM by putting $\lambda_6$ and $\lambda_7$ to be equal to zero, and by making 
$\lambda_5$ and $m_{12}^2$ real. 
Using the definitions of $f_i(v_1,v_2)$ from Eq.\ (\ref{potmin-1}), 
we can write the 
modified minimization condition for one-loop corrected potential $V$ as
\be
f_1(v_1,v_2) + \left. {\frac{\partial V_1}{\partial \phi_1}}\right\vert_{\phi_{c1}=v_1} = 0\,,\ \ \ 
f_2(v_1,v_2) + \left. {\frac{\partial V_1}{\partial \phi_2}}\right\vert_{\phi_{c2}=v_2} = 0 \,,
\label{mod-min}
\ee
where,
\bea
\left.{\frac{\partial V_1}{\partial \phi_1}}\right\vert_{\phi_{c1}=v_1} &=& 
\frac{1}{64\pi^2}\left[ 4m_h^2 F(h,1) 
\left(\ln\frac{m_h^2}{\mu^2} -1\right) + 4m_H^2 F(H,1)
\left(\ln\frac{m_H^2}{\mu^2} -1\right)\right.\nonumber\\
&& + 4 m_A^2 \left(1 - \ln\frac{m_A^2}{\mu^2}\right)\left(v_1 \lambda_{51} 
 - \frac{m_{12}^{2R}}{2v_2}\left(1-\frac{v_2^2}{v_1^2}\right)+\frac{\lambda_{61}v^2}{4v_2}+\frac{\lambda_{61}v_1^2}{2v_2}-\frac{\lambda_{71}v_2v^2}{4v_1^2}+\frac{\lambda_{71}v_2}{2}\right) \nonumber\\
 && + 4 m_{H^\pm}^2 \left(1 - \ln\frac{m_{H^\pm}^2}{\mu^2}\right)\left(v_1 \lambda_{45}
- \frac{m_{12}^{2R}}{v_2}\left(1-\frac{v_2^2}{v_1^2}\right)+\frac{\lambda_{61}v^2}{2v_2}+\frac{\lambda_{61}v_1^2}{v_2}-\frac{\lambda_{71}v_2v^2}{2v_1^2}+\lambda_{71}v_2\right)\nonumber\\
 && -6 g_2^2 m_W^2 v_1 \left(\frac13 - \ln \frac{m_W^2}{\mu^2}\right)
  -3 g^2 m_Z^2 v_1 \left(\frac13 - \ln \frac{m_Z^2}{\mu^2}\right) 
  + 24 Y_b^2 m_b^2 v_1 \left(1 - \ln \frac{m_b^2}{\mu^2}\right) \nonumber\\
  && \left. + 8 Y_{\tau}^2 m_{\tau}^2 v_1 \left(1 - \ln \frac{m_{\tau}^2}{\mu^2}\right)\right]\,
  \label{mod-min-1}
\eea
and
\bea
\left.{\frac{\partial V_1}{\partial \phi_2}}\right\vert_{\phi_{c2}=v_2} &=& 
\frac{1}{64\pi^2}\left[ 4m_h^2 F(h,2)
\left(\ln\frac{m_h^2}{\mu^2} -1\right) + 4m_H^2 F(H,2)
\left(\ln\frac{m_H^2}{\mu^2} -1\right)\right. \nonumber\\
&& + 4 m_A^2 \left(1 - \ln\frac{m_A^2}{\mu^2}\right)\left(v_2 \lambda_{51} 
 - \frac{m_{12}^{2R}}{2v_1}\left(1-\frac{v_1^2}{v_2^2}\right)+\frac{\lambda_{71}v^2}{4v_1}+\frac{\lambda_{71}v_2^2}{2v_1}-\frac{\lambda_{61}v_1v^2}{4v_2^2}+\frac{\lambda_{61}v_1}{2}\right) \nonumber\\
 && + 4 m_{H^\pm}^2 \left(1 - \ln\frac{m_{H^\pm}^2}{\mu^2}\right)\left(v_2 \lambda_{45}
- \frac{m_{12}^{2R}}{v_1}\left(1-\frac{v_1^2}{v_2^2}\right)+\frac{\lambda_{71}v^2}{2v_1}+\frac{\lambda_{71}v_2^2}{v_1}-\frac{\lambda_{61}v_1v^2}{2v_2^2}+\lambda_{61}v_1\right)\nonumber\\
 && \left. -6 g_2^2 m_W^2 v_2 \left(\frac13 - \ln \frac{m_W^2}{\mu^2}\right)
  -3 g^2 m_Z^2 v_2 \left(\frac13 - \ln \frac{m_Z^2}{\mu^2}\right) 
  + 24 Y_t^2 m_t^2 v_2 \left(1 - \ln \frac{m_t^2}{\mu^2}\right)\right]\,,
  \label{mod-min-2}
\eea
with
$\lambda_{45} = \lambda_4 + \lambda_{51} $ and
$ g = \sqrt{g_1^2 + g_2^2} $, $g_1$ and $g_2$ being the $U(1)_Y$ and $SU(2)_L$ gauge couplings. 
Yukawa couplings for $t$, $b$, and $\tau$ are 
denoted by $Y_t$, $Y_b$ and $Y_{\tau}$ respectively.
The $F$-functions have been defined in Appendix A. 
Note that these expressions 
are valid only if we tune $\mu$ to keep the position of the EW minimum of the one-loop corrected effective potential $V$,
defined in Eq.\ (\ref{V_eff}), unchanged
with respect to its position in the tree level potential.
One also needs to check the second derivatives to ensure that the extremum is a local minimum:
\be
\left. \frac{\partial^2 V}{\partial \phi_1^2}\right\vert_{\phi_{c1}=v_1} > 0 \,,\ \ \ 
\left. \frac{\partial^2 V}{\partial \phi_2^2}\right\vert_{\phi_{c2}=v_2} > 0 \,.
\label{mod-min-3}
\ee
The expressions for second derivatives are also given in Appendix A.
Note the absence of the Goldstone bosons in Eqs.\ (\ref{mod-min},\ref{mod-min-1},\ref{mod-min-2}) 
because of our choice of $\mu$ which keeps them massless.

\section{Analysis and Results}
\subsection{2HDM at tree-level} 

From a random scan over the parameter space spanning over $7\times 10^8$ different choices of model
parameters, we generate a number of models for which the following 
conditions are satisfied. Our analysis includes the canonical 2HDM with $Z_2$ symmetry, and 2HDM without 
$Z_2$ with both real and complex parameters. 

\begin{itemize}
 \item The potential has to be stable at all scales before it either blows up (due to one or more couplings
 hitting the Landau pole) or becomes unbounded from below. 
 
 \item The dimensionless couplings must remain perturbative over the entire range of validity of the theory, except maybe at 
 the very end where they approach the Landau pole. 
 
 \item  There should be one minimum of the scalar potential for which $v = \sqrt{v_1^2 + v_2^2} = 246$ GeV. This we will
 call the EW vacuum. This, in fact, acts as the tightest constraint and rules out the largest chunk of randomly generated 
 models. We focus only on those models that allow another local minimum apart from the EW vacuum. Both tree-level
 minimization conditions, for $v_1$ and $v_2$,  should be satisfied in both the vacua. 

 \item Other constraints like that on the charged Higgs boson mass coming from $b\to s\gamma$ are satisfied. We use
 $m_{H^\pm} > 316$ GeV. While not all
 constraints are valid for all 2HDMs, we focus only on the Type II 2HDM. 
 
 \item The 125 GeV resonance found at the Large Hadron Collider must have properties close to that of the SM Higgs
 boson. In other words, the alignment limit should be maintained. We have kept $\vert\alpha-\beta\vert$ to be between 
 $0.9\pi/2$ and $1.1\pi/2$, noting that this is a rather conservative limit. 

\end{itemize}

The ranges for our scan is as follows:   
\be
0.0 \leq \lambda_1, \lambda_2 \leq 1.0\,, \ \ 
-1.0 \leq \lambda_3, \lambda_4, \lambda_5\leq 1.0\,, \ \
m^2_{12} \leq 4\times 10^6~~{\rm GeV}^2\,,\ \ 
1.0 \leq \tan\beta \leq 50.0\,,
\ee
where the parameters are taken to be at the EW scale. 

Apart from this, for the $Z_2$-violating cases, we have taken
\be
-1.0 \leq  \lambda_{6}, \lambda_{7} \leq 1.0 \ \ ({\rm real~couplings})\,,\ \ \ 
-1.0 \leq \lambda_{51}, \lambda_{52}, \lambda_{61}, \lambda_{62}, \lambda_{71}, \lambda_{72} \leq 1.0\ \ 
({\rm complex~couplings})\,.
\ee
Also, for this case the scan is made on $m_{12}^2 = m_{12}^{2R} + i m_{12}^{2I}$, as the phase in $m_{12}^2$ is
fixed by Eq.\ (\ref{zero-gold}). 

We find a few common characteristics for these models that allow two minima. They are: \\
(i) The EW vacuum is always deeper than the other vacuum. This happens mostly because of the imposition of the 
experimental constraints. Thus, even with the introduction of $Z_2$-breaking parameters, there is no chance of tunneling 
to the other minimum, at least at the tree-level. This reinforces the conclusions obtained by the authors of Ref.\ \cite{Barroso:2013awa}. \\
(ii) If both the vacua are normal, there is no other minimum that breaks charge conservation or $CP$. This, again, is in tune 
of what the authors of Refs.\ \cite {Ferreira:2004yd,Ivanov:2007de} found.


\subsection{2HDM at one-loop level}


We would now like to perform the same analysis on the one-loop corrected potential on those models that satisfy  
the initial constraints and show the presence of a double minima at the tree-level. The regularization scale $\mu$ is so chosen 
as to make 
$\phi_{c1}=v_1$ and $\phi_{c2}=v_2$. This is, of course, a rather restrictive choice, but keeps all the masses as well as 
$\tan\beta$ unchanged at the EW minimum even after the one-loop corrections are implemented. 

In a generic 2HDM where both CP-even neutral fields can get nonzero VEV, this is a tricky job, and mathematically much more 
complicated than the cases of only SM, or SM extended by a gauge singlet scalar, or the inert doublet models where one VEV is 
always zero. 
The complication is further enhanced by the fact that at the tree-level, there are two minima of the potential.

What we do is the following. We fix $\phi_{c1}=v_1$ and tune $\mu$ in such a way that $\phi_{c2}$ 
coincides with $v_2$. This keeps the 
position of the EW minimum invariant but changes the depth. We could have done this the other way round too, namely, keeping 
$\phi_{c2}=v_2$ and adjusting $\mu$ to make $\phi_{c1}$ coincide with the tree level value; 
however, we prefer the first approach as the contribution of $v_2$ is larger in $v$ for
$\tan\beta > 1$. We then use the same $\mu$ but fix the classical minimum for $\phi_1$ at 
\footnote{We use primes to denote the corresponding quantities in the second minimum.} 
$\phi'_{c1} = v'_1$. It so happens that $\phi'_{c2}$, in all cases, lies close to 
$v'_2$; the coincidence is not as exact as the EW vacuum, but this being the shallower secondary minimum, we will not be so much 
bothered about its exact position.  Thus, we compare $V(\mu,\phi_{c1}=v_1,
\phi_{c2}=v_2)$ with $V'(\mu,\phi'_{c1}=v'_1,\phi'_{c2}\approx v'_2)$. We will also show the effect of varying the scale $\mu$ on the 
classical field values.

What we observe is that the conclusions drawn from a tree-level analysis is more or less unchanged; the deeper minimum 
remains deeper. This is not entirely unexpected, as the one-loop corrections are only a small effect that cannot overcome 
the difference in depths of these two minima. For $\tan\beta \gg 1$, $\phi_1$ direction is almost flat, so we show our results 
in the constant-$\phi_1$ direction, varying the potential $V$ with $\phi_2$. 

Our results are discussed in more detail later for some typical benchmark points, but for both the 
cases that we consider ($Z_2$ conserving and breaking), we never found the second minimum becoming 
deeper than the EW one after the one-loop corrections. Thus, if the EW vacuum is the deeper one at 
tree-level, it remains so after the radiative corrections; there is no chance of developing a deeper 
vacuum and tunnelling into it. In fact, if $\mu$ is tuned in such a way that the EW vacuum is not shifted from its tree-level position, 
{\em it always gets deeper by the one-loop corrections}; $V_1$ at the EW vacuum is always negative. 

\subsection{One-loop corrected 2HDM with $Z_2$ symmetry} 

In Table \ref{tab:tab-bench}, we show three benchmark points for the $Z_2$-conserving Type-II 2HDM, characterized by 
small, medium, and large values of $\tan\beta (=v_2/v_1)$ respectively. All these models show the existence of a second 
and shallower minimum compared to the EW one. These three models, namely, $Z_2C1$, $Z_2C2$, and $Z_2C3$ are 
valid up to $3.8\times 10^7$ GeV, $2.8\times 10^{11}$ GeV, and the Planck scale respectively; for the first two models,
at least one of the couplings become nonperturbative at the validity scale and the model soon hits the Landau pole 
thereafter. Obviously, the comparatively low validity range for the first benchmark can be ascribed to the relatively large
quadratic couplings to start with at the EW scale.  The evolutions are checked with one-loop renormalization group equations 
for the Type II 2HDM \cite{Branco:2011iw,Chakraborty:2014oma}.

\begin{table}[htbp]
\begin{center}
\begin{tabular}{||c||c|c|c||}
\hline
 & \multicolumn{3}{c||} {Benchmark} \\
 \hline
&&&\\
Parameter                     & $Z_2C1$ & $Z_2C2$ & $Z_2C3$ \\
&&&\\
\hline

$\lambda_1$                 & 0.413      & 0.642 & 0.068  \\
$\lambda_2$                 & 0.842      & 0.328 & 0.260  \\
$\lambda_3$                 & $-0.265$ & 0.065 & 0.132  \\
$\lambda_4$                 & $-0.720$ & $-0.365$ & $-0.489$  \\
$\lambda_5$                 & $-0.929$ & $-0.786$ & $-0.498$  \\
$m_{11}^2$ (GeV$^2$) & $2.3\times 10^6$ & $6.2\times 10^5$ & $2.61\times 10^5$ \\
$m_{22}^2$ (GeV$^2$) & $2.8\times 10^5$ & $5.2\times 10^3$ & $ -7.77\times 10^3$ \\
$m_{12}^2$ (GeV$^2$)& $8.10\times 10^5$ & $9.11\times 10^4$ & $4.84\times 10^3$ \\
$v_1$ (GeV)                & 83.76 & 37.65 & 5.05 \\
$v_2$ (GeV)                & 231.30 & 243.11 & 245.95 \\
$v'_1$ (GeV)                & 442.46 & 452.48& 399.18  \\
$v'_2$ (GeV)                & 872.82 & 953.66 & 775.94  \\
\hline

\end{tabular}
\end{center}
\caption{Benchmark points for $Z_2$-conserving Type-II 2HDM. 
}
\label{tab:tab-bench}
\end{table}


The positions of the 
one-loop corrected minima and the corresponding regularization scales are shown in Table
\ref{tab:tab-result}. The benchmarks are chosen scanning the range of $\tan\beta$ as well as
the other parameters.

\begin{table}[htbp]
\begin{center}
\begin{tabular}{||c|c|c|c|c|c|c||}
\hline
    &  $\tan\beta$ & $\phi_{c1}$ & $\phi_{c2}$ & $\mu$ & $\phi'_{c1}$ & $\phi'_{c2}$ \\
\hline
$Z_2C1$ & 2.76 & 83.76 & 231.30 & 976.0 & 442.46 & 846.7 \\
$Z_2C2$ & 6.46 & 37.65 & 243.11 & 492.5 & 452.48 & 932.6 \\
$Z_2C3$ & 48.7 & 5.05 & 245.95 & 335.0 & 953.66 & 774.6 \\
\hline
\end{tabular}
\end{center}
\caption{The one-loop corrected minima (all quantities except tan$\beta$ are in GeV) for the three benchmarks. 
Note the tiny shift of $\phi'_{c2}$ from $v'_2$.
}
\label{tab:tab-result}
\end{table}

\begin{figure}[htbp]
\begin{center}
\hspace*{-5mm}\includegraphics[height=5cm,width=7cm]{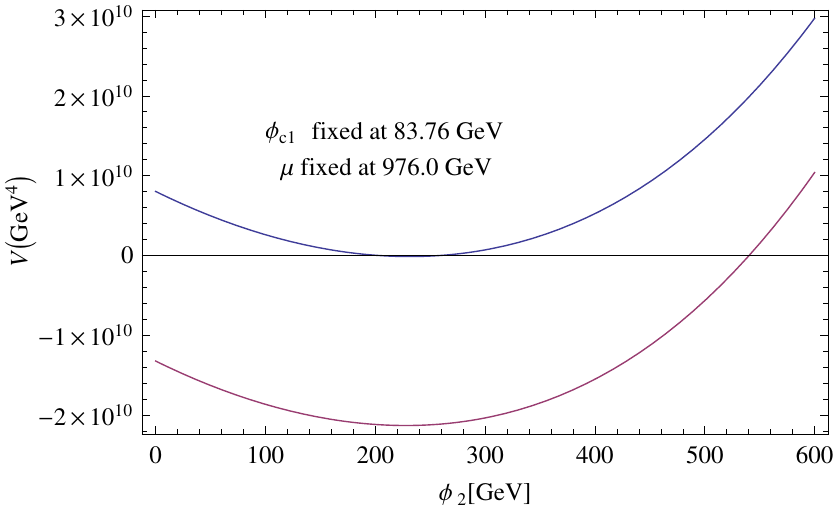} \hspace{2cm}
\includegraphics[height=5cm,width=7cm]{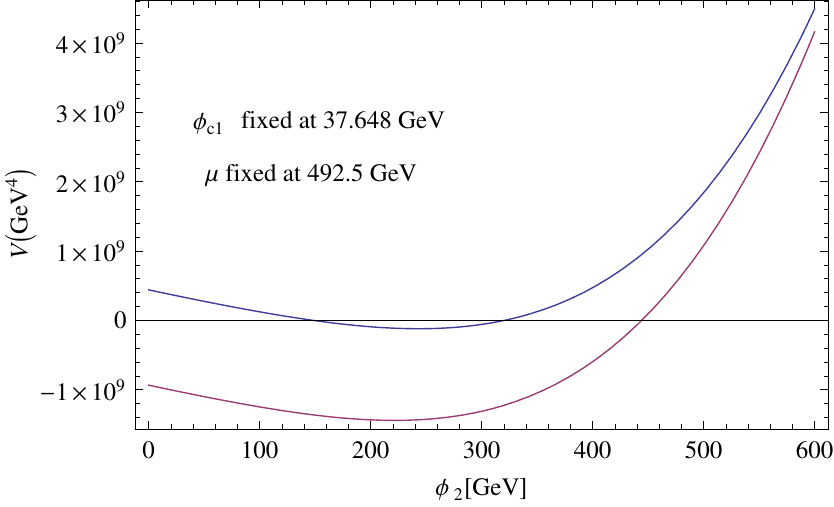}\\[5mm]
 \includegraphics[height=5cm,width=7cm]{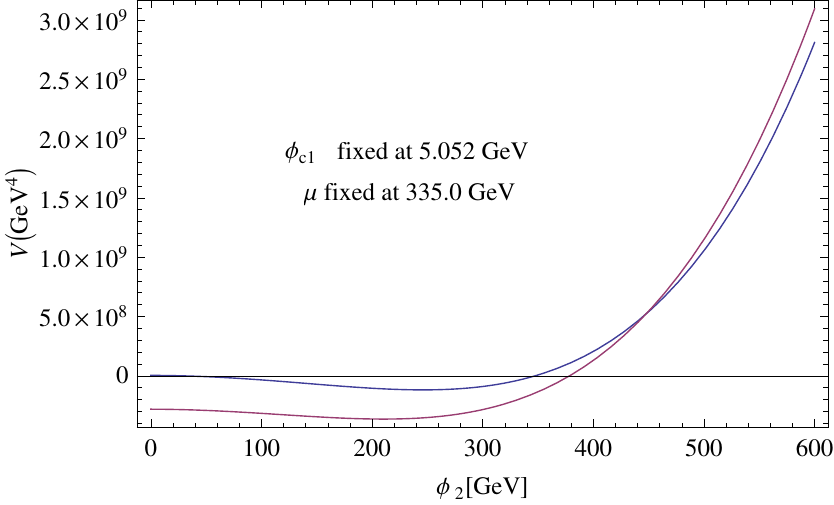}
  \end{center}
  \caption{The plots of the tree-level and one-loop corrected potential, with the section taken at a fixed value of
  $\phi_{c1}$ as indicated on the plots and in Table \ref{tab:tab-result}. 
  The upper panel plots are for benchmarks $Z_2C1$ (L) and $Z_2C2$ (R),
  while the lower panel plot is for $Z_2C3$. In every plot, the upper curve (blue) denotes the tree-level potential profile, 
  while the lower one (red) is for the one-loop corrected potential.}
 \label{fig:pot-z2symm}
\end{figure}

The potential profiles are shown in Figure \ref{fig:pot-z2symm}. These are drawn as a section of the actual three-dimensional
plots, for fixed values of $\phi_{c1}$. That is why the second minimum is not apparent; it occurs at a different value of $\phi_{c1}$. 
Also, the tree-level potential does not go to zero as $\phi_{c2}\to 0$, unless $\phi_{c1}$ is tiny, as in $Z_2C3$. 
Note that the one-loop corrected potential is always deeper than the tree-level potential at the EW minimum; Table 
\ref{tab:tab-result} shows that the second minimum $\phi'_{c2}$ almost coincides with $v'_2$.

 Only if $\phi_{c1}$ is small, like in $Z_2C3$, the $\phi_1$ direction can be approximated by a flat direction. The flatness is really 
 impressive: for a 10\% (1\%) change in $\phi_{c1}$, the potential changes only by $0.02\%$ ($2.5\times 10^{-4}\%$). 
 However, we have not found any case where the one-loop corrections remove the second minimum. 
 
\begin{figure}[htbp]
\hspace*{-5mm}
\includegraphics[height=5cm,width=7cm]{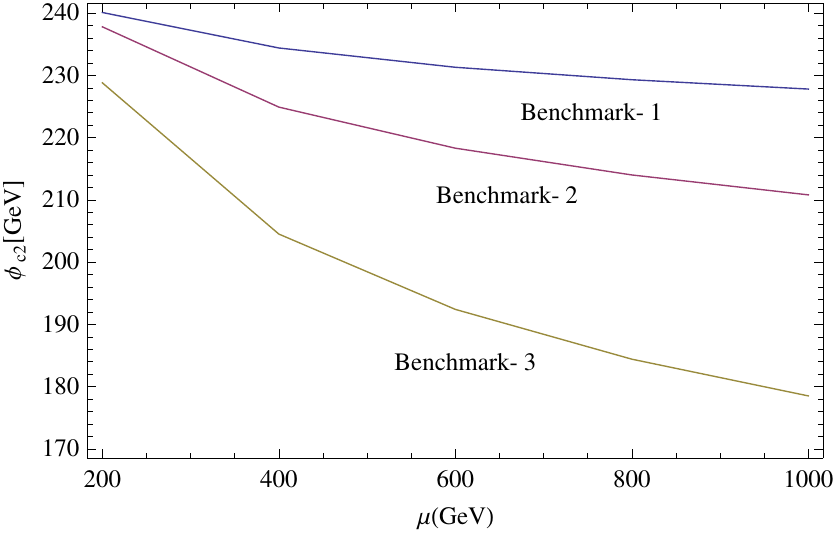}  \hspace*{2cm}
\includegraphics[height=5cm,width=7cm]{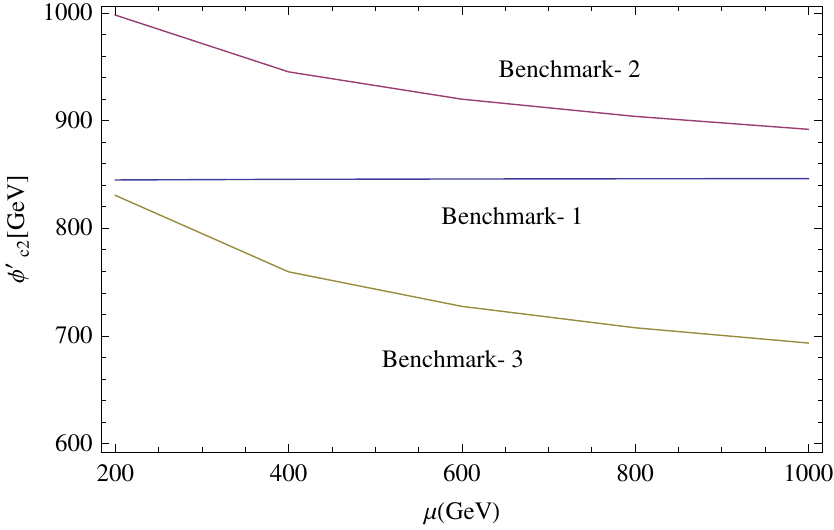}
\caption{Variation of $\phi_{c2}$ (L) and $\phi'_{c2}$ (R) with $\mu$ for $Z_2C1$ (blue), $Z_2C2$ (red), and $Z_2C3$ 
(golden).}
\label{fig:varz2symm}
\end{figure}

If we keep $\phi_{c1}=v_1$ or $v'_1$, and play with $\mu$ as a free parameter, $\phi_{c2}$ and  $\phi'_{c2}$ changes.
In Fig. \ref{fig:varz2symm}, we show how  $\phi_{c2}$ and $\phi'_{c2}$ change with $\mu$ for the three benchmarks.

 \subsection{One-loop corrected 2HDM without $Z_2$ symmetry} 
 
\begin{figure}[htbp!]
\begin{center}
\hspace*{-5mm}\includegraphics[height=5cm,width=7cm]{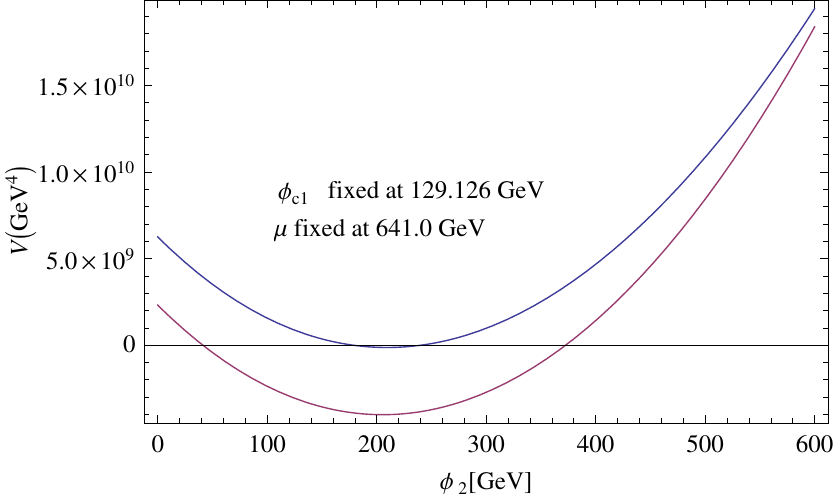}\hspace*{2cm}
\includegraphics[height=5cm,width=7cm]{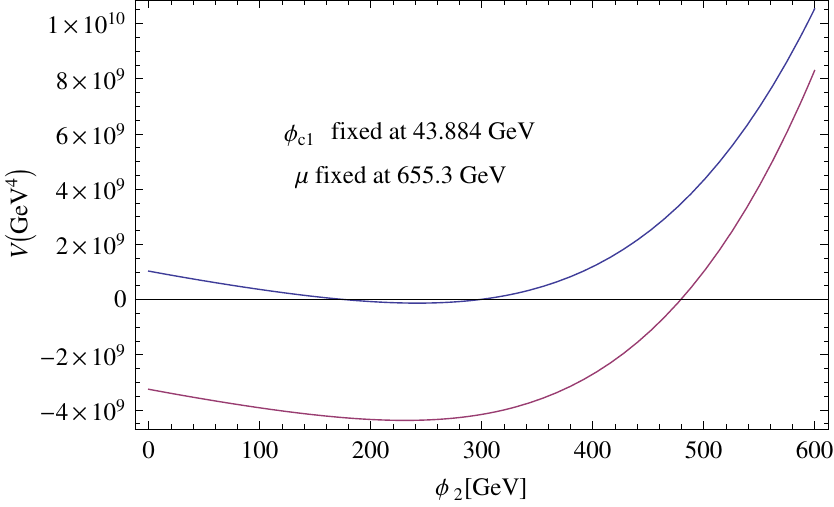}\\[5mm]
 \includegraphics[height=5cm,width=7cm]{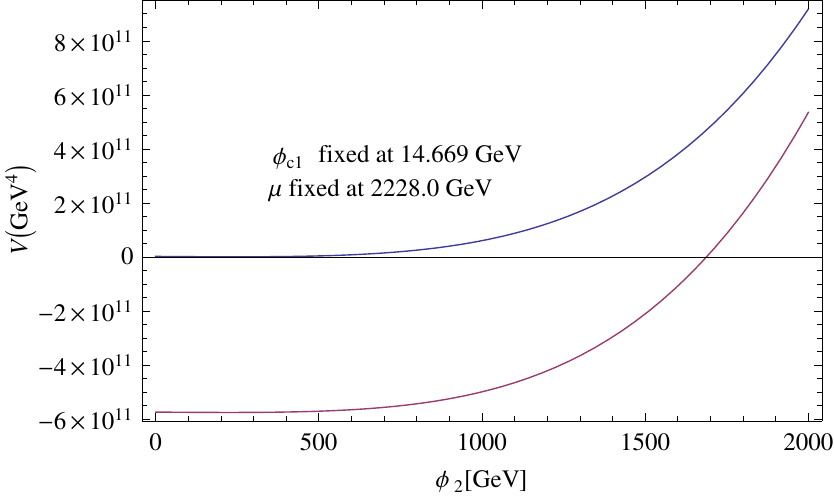}
  \end{center}
 \caption{The plots of the tree-level and one-loop corrected potential, with the section taken at a fixed value of
  $\phi_{c1}$ as indicated on the plots. 
  The upper panel plots are for benchmarks $Z_2V1$ (L) and $Z_2V2$ (R),
  while the lower panel plot is for $Z_2V3$. In every plot, the upper curve (blue) denotes the tree-level potential profile, 
  while the lower one (red) is for the one-loop corrected potential.}
 \label{fig:pot-z2asymm}
\end{figure}

The analysis is analogous to what was performed for the $Z_2$-symmetric case. The potential profiles are shown in 
Fig.\ \ref{fig:pot-z2asymm}. For the three benchmark points $Z_2V1$, $Z_2V2$, and $Z_2V3$, the $\mu$-values are 
fixed at 641.0 GeV, 655.3 GeV, and 2228 GeV respectively. While $\phi_{c1}=v_1$, $\phi_{c2}=v_2$, and $\phi'_{c1}=v'_1$
were ensured, the $\phi'_{c2}$ values were quite close to that of $v'_2$; they are at 506.2 GeV, 708.9 GeV, and 1783 GeV
respectively. 

\begin{table}[htbp]
\begin{center}
\begin{tabular}{||c||c|c|c||}
\hline
 & \multicolumn{3}{c||} {Benchmark} \\
 \hline
&&&\\
Parameter                     & $Z_2V1$ & $Z_2V2$ & $Z_2V3$ \\
&&&\\
\hline

$\lambda_1$                 & 0.656      & 0.342 & 0.497  \\
$\lambda_2$                 & 0.188      & 0.928 & 0.456  \\
$\lambda_3$                 & $0.836$ & 0.998 & $0.089$  \\
$\lambda_4$                 & $0.659$ & $0.375$ & $0.598$  \\
$\lambda_{51}$                 & $0.764$ & $-0.956$ & $-0.533$  \\
$\lambda_{52}$                 & $0.163$ & $0.666$ & $0.923$  \\
$\lambda_{61}$                 & $0.633$ & $0.735$ & $0.680$  \\
$\lambda_{62}$                 & $0.209$ & $0.619$ & $-0.929$  \\
$\lambda_{71}$                 & $-0.820$ & $-0.911$ & $-0.810$  \\
$\lambda_{72}$                 & $-0.0016$ & $0.709$ & $0.774$  \\
$m_{11}^2$ (GeV$^2$) & $7.47\times 10^5$ & $1.06\times 10^6$ & $1.34\times 10^7$ \\
$m_{22}^2$ (GeV$^2$) & $3.1\times 10^5$ & $1.78\times 10^4$ & $ 3.71\times 10^4$ \\
$m_{12}^{2R}$ (GeV$^2$)& $4.92\times 10^5$ & $1.71\times 10^5$ & $7.78\times 10^5$ \\
$m_{12}^{2I}$ (GeV$^2$)& $3.92\times 10^3$ & $2.49\times 10^4$ & $2.49\times 10^4$ \\
$v_1$ (GeV)                & 129.13 & 43.89 & 14.67 \\
$v_2$ (GeV)                & 209.39 & 242.05 & 245.56 \\
$v'_1$ (GeV)                & 295.20 & 257.06& 355.55  \\
$v'_2$ (GeV)                & 631.54 & 799.49 & 2064.3  \\
\hline

\end{tabular}
\end{center}
\caption{Benchmark points for $Z_2$-violating Type-II 2HDM. Note that $\lambda_i =
\lambda_{i1} + i\lambda_{i2}$ for $i=5,6,7$. Also, $m_{12}^2 = m_{12}^{2R} + 
im_{12}^{2I}$. 
}
\label{tab:tab-bench2}
\end{table}

  \begin{figure}[htbp]
\hspace*{-5mm}
\includegraphics[height=5cm,width=7cm]{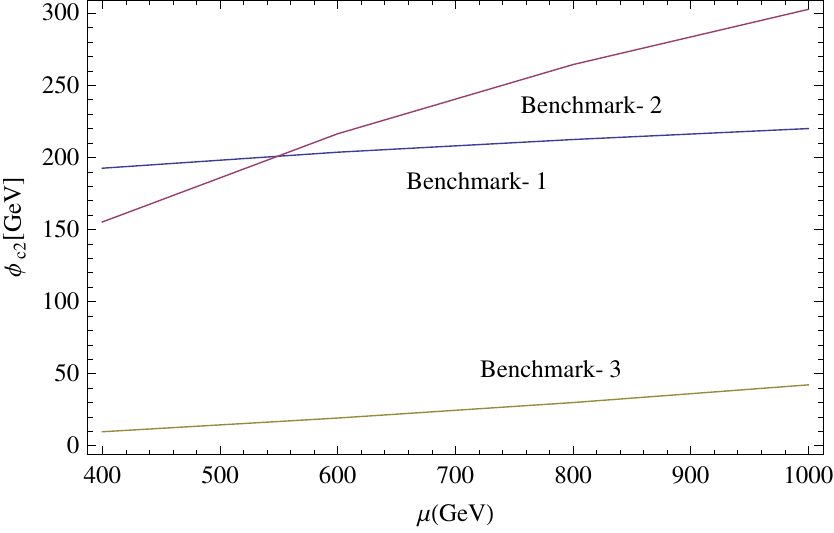} \hspace*{2cm}
\includegraphics[height=5cm,width=7cm]{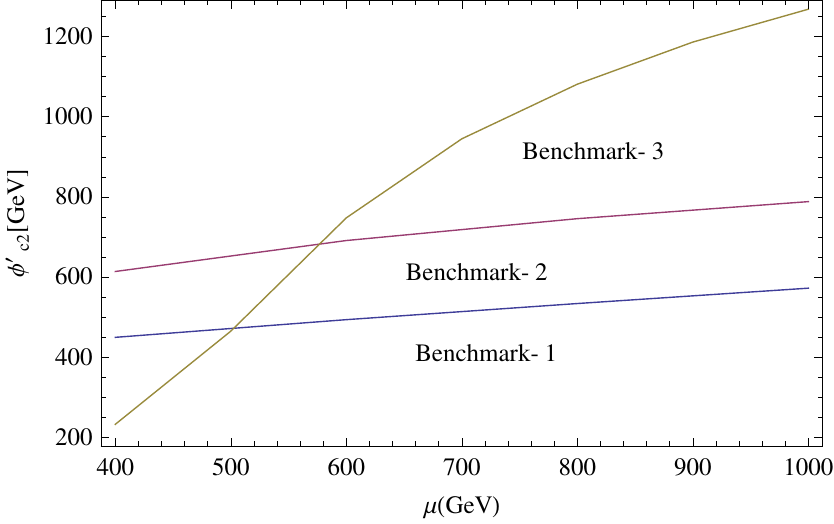}
\caption{Variation of $\phi_{c2}$ (L) and $\phi'_{c2}$ (R) with $\mu$ for $Z_2V1$ (blue), $Z_2V2$ (red), and $Z_2V3$ 
(golden).}
\label{fig:varz2asymm}
\end{figure}

 Just like the $Z_2$-conserving case, we choose three benchmarks for the case where the potential violates the $Z_2$ and 
 also involves complex parameters. These benchmarks are shown in Table \ref{tab:tab-bench2}.

The flatness of the potential in the $\phi_1$ direction is again most manifest for $Z_2V3$, the benchmark with 
lowest $v_1$. At the same time, the EW minimum always gets deeper with the one-loop corrected potential, 
and there is no qualitative change from the tree-level result. Because of the comparatively large values of the 
couplings, all the three models tend to hit the Landau pole much before the Planck scale, namely, at 45.8 TeV, 
4.8 TeV, and 216.7 TeV. This is expected because $Z_2V2$ starts with larger values of the quartic couplings at the 
EW scale. In general, for the double-minima case when $Z_2$ is broken, most of the couplings at the EW scale 
have to be large to start with and thus such class of models are not stable beyond a few hundreds of TeV, which
may be contrasted with the $Z_2$-conserving double-minima models.

In Fig. \ref{fig:varz2asymm}, we show how  $\phi_{c2}$ and $\phi'_{c2}$ changes with $\mu$ for the three benchmarks.

   \section{Summary}

In this paper we have tried to investigate the nature of the potential of Type II 2HDM, breaking the $Z_2$ symmetry 
either softly or through dimension-4 operators. There are some known results for the $Z_2$-symmetric 2HDM at the 
tree level. Our goal was to investigate how far these conslusions are reliable if one (i) breaks the $Z_2$ symmetry 
at the tree level with operators with real or complex couplings, (ii) does a one-loop correction on the 
potential. 

For the first part, we find that the introduction of $Z_2$ breaking does not change the conclusions qualitatively: 
the scalar potential can accommodate at most two local minima, both of which have to be normal. If there is a normal 
minimum, there cannot be a charge-breaking or $CP$ violating minimum of the potential. The LHC data highly disfavours 
those models where the second minimum is deeper than the EW minimum, possibly making the EW vacuum an unstable or
metastable one. 

For one-loop corrections, we use a regularization scale that keeps the position of the EW minimum invariant, changing 
only its depth. When we focus on models with two minima, this prescription keeps the position of the second 
minima almost unchanged too. The one-loop corrections cannot change the 
relative depths of the minima, {\em i.e.}, the EW minimum still remains deeper after the correction, ruling out the 
possibility of a metastable vacuum. The conclusions are identical for $Z_2$ symmetric and $Z_2$ breaking 2HDM. 
While the conclusions were drawn for Type II 2HDM, the results are qualitatively the same for other 2HDMs too, as 
the only change comes from the Yukawa couplings that enter the one-loop corrections.

The stability of the 2HDM at higher energy scales is a complex issue because of more fields and couplings. One
needs the stability conditions to be valid at all scales and the couplings to remain perturbative for calculability 
(or, at least, not blow up). A general tendency that can intuitively be deduced from the RG equations is that higher 
values of quartic couplings at the EW scale tend to pull down the range of validity of the theory, which means 
that some other ultraviolet complete theory takes hold beyond that range. However, a large part of the parameter
space is still compatible with the stability up to the Planck scale for the $Z_2$-conserving class of models. 
For the $Z_2$-violating class, the existence of two minima generally forces some of the quartic couplings to be large
at the EW scale and hence such models cease to be valid beyond a few hundreds of TeV at the most; for smaller 
couplings, one gets the single-minimum models.

\centerline{\bf{Acknowledgements}}

The authors acknowledge Nabarun Chakrabarty and Dipankar Das for several helpful discussions. 
I.C.\ acknowledges the Council for Scientific and Industrial Research, Government of India, for a 
research fellowship. A.K.\ acknowledges the Department of Science and Technology, Government of  
India, and the Council for Scientific and Industrial Research, Government of India, for support 
through research grants. 


 \begin{appendix}
\numberwithin{equation}{section}

\section{Expressions for first and second derivatives}

We use the following shorthand notations:
\be
F(\alpha,i) =  \left. m_\alpha \left( \frac{\partial m_{\alpha}}{\partial \phi_i}\right) 
\right\vert_{\phi_{ci}=v_i}\,,\ \ \ 
G(\alpha,i) = \left[\left( \frac{\partial m_\alpha}{\partial \phi_i} \right)^2 + 
m_\alpha \left( \frac{\partial^2 m_\alpha}{\partial \phi_i^2} \right) \right]_{\phi_{ci}=v_i}\,.
\ee

\bea
F(h,1) &=& \frac14\left(2\lambda_1 v_1 + \frac{m_{12}^{2R}}{v_1^2v_2}\left(v_1^2-v_2^2\right)+\frac{3\lambda_{67}v_2}{2}+\frac{\lambda_{71}v_2^3}{2v_1^2}
-\frac{3\lambda_{61}v_1^2}{2v_2}\right) \nonumber\\
 && -\frac{1}{4\sqrt{a}}\left(\lambda_1 v_1^2 - \lambda_2 v_2^2 -\frac{m_{12}^{2R}}{v_1 v_2} \left(v_1^2-v_2^2\right)+\frac{3v_1v_2\left(\lambda_{61}-\lambda_{71}\right)}{2}
 -\frac{\lambda_{71}v_2^3}{2v_1}+\frac{\lambda_{61}v_1^3}{2v_2}\right)\times \nonumber\\
 &&\left(\lambda_1 v_1 -\frac{m_{12}^{2R}}{v_2}
 + \left(v_1^2- v_2^2\right)\frac{m_{12}^{2R}}{2 v_1^2 v_2}+\frac{3v_2\left(\lambda_{61}-\lambda_{71}\right)}{4}+\frac{\lambda_{71}v_2^3}{4v_1^2}+\frac{3\lambda_{61}v_1^2}{4v_2}\right) \nonumber\\
 && -\frac{1}{2 \sqrt{a}}\left(-m_{12}^{2R} + 
 \lambda_{345} v_1 v_2+\frac{3\lambda_{61}v_1^2}{2}+\frac{3\lambda_{71}v_2^2}{2}\right)\times 
 \left(\lambda_{345} v_2+3\lambda_{61}v_1\right)\,,\nonumber\\
 F(h,2) &=& \frac14\left(2\lambda_2 v_2 + \frac{m_{12}^{2R}}{v_2^2v_1}\left(v_2^2-v_1^2\right)+\frac{3\lambda_{67}v_1}{2}-\frac{3\lambda_{71}v_2^2}{2v_1}
 +\frac{\lambda_{61}v_1^3}{2v_2^2}\right) \nonumber\\
 && -\frac{1}{4\sqrt{a}}\left(\lambda_1 v_1^2 - \lambda_2 v_2^2 -\frac{m_{12}^{2R}}{v_1 v_2} \left(v_1^2-v_2^2\right)+\frac{3v_1v_2\left(\lambda_{61}-\lambda_{71}\right)}{2}
 -\frac{\lambda_{71}v_2^3}{2v_1}+\frac{\lambda_{61}v_1^3}{2v_2}\right)\times\nonumber\\
 &&\left(-\lambda_2 v_2 +\frac{m_{12}^{2R}}{v_1}
 + \left(v_1^2- v_2^2\right)\frac{m_{12}^{2R}}{2 v_2^2 v_1}+\frac{3v_1\left(\lambda_{61}-\lambda_{71}\right)}{4}-\frac{3\lambda_{71}v_2^2}{4v_1}-\frac{\lambda_{61}v_1^3}{4v_2^2}\right) \nonumber\\
 && -\frac{1}{2 \sqrt{a}}\left(-m_{12}^{2R} + \lambda_{345} v_1 v_2+\frac{3\lambda_{61}v_1^2}{2}+\frac{3\lambda_{71}v_2^2}{2}\right)\times\left(\lambda_{345} v_1+3\lambda_{71}v_2\right)\,,\nonumber\\
F(H,1) &=& \frac14\left(2\lambda_1 v_1 + \frac{m_{12}^{2R}}{v_1^2v_2}\left(v_1^2-v_2^2\right)+\frac{3\lambda_{67}v_2}{2}+\frac{\lambda_{71}v_2^3}{2v_1^2}
-\frac{3\lambda_{61}v_1^2}{2v_2}\right) \nonumber\\
 && +\frac{1}{4\sqrt{a}}\left(\lambda_1 v_1^2 - \lambda_2 v_2^2 -\frac{m_{12}^{2R}}{v_1 v_2} \left(v_1^2-v_2^2\right)+\frac{3v_1v_2\left(\lambda_{61}-\lambda_{71}\right)}{2}
 -\frac{\lambda_{71}v_2^3}{2v_1}+\frac{\lambda_{61}v_1^3}{2v_2}\right)\times \nonumber\\
 &&\left(\lambda_1 v_1 -\frac{m_{12}^{2R}}{v_2}
 + \left(v_1^2- v_2^2\right)\frac{m_{12}^{2R}}{2 v_1^2 v_2}+\frac{3v_2\left(\lambda_{61}-\lambda_{71}\right)}{4}+\frac{\lambda_{71}v_2^3}{4v_1^2}+\frac{3\lambda_{61}v_1^2}{4v_2}\right) \nonumber\\
 && +\frac{1}{2 \sqrt{a}}\left(-m_{12}^{2R} + \lambda_{345} v_1 v_2+\frac{3\lambda_{61}v_1^2}{2}+\frac{3\lambda_{71}v_2^2}{2}\right)\times \left(\lambda_{345} v_2+3\lambda_{61}v_1\right)\,,\nonumber\\
 F(H,2) &=& \frac14\left(2\lambda_2 v_2 + \frac{m_{12}^{2R}}{v_2^2v_1}\left(v_2^2-v_1^2\right)+\frac{3\lambda_{67}v_1}{2}-\frac{3\lambda_{71}v_2^2}{2v_1}
 +\frac{\lambda_{61}v_1^3}{2v_2^2}\right) \nonumber\\
 && +\frac{1}{4\sqrt{a}}\left(\lambda_1 v_1^2 - \lambda_2 v_2^2 -\frac{m_{12}^{2R}}{v_1 v_2} \left(v_1^2-v_2^2\right)+\frac{3v_1v_2\left(\lambda_{61}-\lambda_{71}\right)}{2}
 -\frac{\lambda_{71}v_2^3}{2v_1}+\frac{\lambda_{61}v_1^3}{2v_2}\right)\times\nonumber\\
 &&\left(-\lambda_2 v_2 +\frac{m_{12}^{2R}}{v_1}
 + \left(v_1^2- v_2^2\right)\frac{m_{12}^{2R}}{2 v_2^2 v_1}+\frac{3v_1\left(\lambda_{61}-\lambda_{71}\right)}{4}-\frac{3\lambda_{71}v_2^2}{4v_1}-\frac{\lambda_{61}v_1^3}{4v_2^2}\right) \nonumber\\
 && +\frac{1}{2 \sqrt{a}}\left(-m_{12}^{2R} + \lambda_{345} v_1 v_2+\frac{3\lambda_{61}v_1^2}{2}+\frac{3\lambda_{71}v_2^2}{2}\right)\times\left(\lambda_{345} v_1+3\lambda_{71}v_2\right)\,,\nonumber\\
\eea
where 
\bea
\lambda_{67}&=& \lambda_{61} + \lambda_{71} \,, \ \ 
\lambda_{345} = \lambda_3 + \lambda_4 + \lambda_{51}\,, \nonumber\\
a &=& \left(\frac{\left(\lambda_1v_1^2-\lambda_2v_2^2\right)}{2}+\frac{3v_1v_2\left(\lambda_{61}-\lambda_{71}\right)}{4}-\frac{\lambda_{71}v_2^3}{4v_1}+\frac{\lambda_{61}v_1^3}{4v_2}-
\frac{m_{12}^{2R}\left(v_1^2-v_2^2\right)}{v_1v_2}\right)^2 \nonumber\\
&&+\left(-m_{12}^{2R}+\lambda_{345}v_1v_2+\frac{3\lambda_{61}v_1^2}{2}+\frac{3\lambda_{71}v_2^2}{2}\right)^2\,.
\eea

\bea
 \left. \frac{\partial^2 V_1}{\partial \phi_1^2}\right\vert_{\phi_{c1}=v_1} 
 &=& 
 \frac{1}{64\pi^2} \left[ 4 m_A^2 \left(\frac{m_{12}^{2R} v_2}{v_1^3} - \lambda_{51}-\frac{3\lambda_{61}v_1}{2v_2}
+\frac{\lambda_{71}v_2}{2v_1}
-\frac{\lambda_{71}v_2v^2}{2v_1^3}\right)\left(\ln \frac{m_A^2}{\mu^2} - 1\right)\right. \nonumber\\
&& + 8 \left(\frac{m_{12}^{2R}}{2v_2}\left(1-\frac{v_2^2}{v_1^2}\right)-
v_1 \lambda_{51}-\frac{\lambda_{61}v^2}{4v_2}-\frac{\lambda_{61}v_1^2}{2v_2}-
\frac{\lambda_{71}v_2}{2}+\frac{\lambda_{71}v_2v^2}{4v_1^2}\right)^2 \ln \frac{m_A^2}{\mu^2} \nonumber\\
 && + 8 m_{H^\pm}^2 \left(\frac{m_{12}^{2R} v_2}{v_1^3} - \frac12\lambda_{45}-\frac{3\lambda_{61}v_1}{2v_2}
+\frac{\lambda_{71}v_2}{2v_1}
-\frac{\lambda_{71}v_2v^2}{2v_1^3}\right)\left(\ln \frac{m_{H^\pm}^2}{\mu^2} - 1\right) \nonumber\\
&&+ 4  \left(\frac{m_{12}^{2R}}{v_2}\left(1-\frac{v_2^2}{v_1^2}\right)-v_1 \lambda_{45}-\frac{\lambda_{61}v_1^2}{v_2}-\frac{\lambda_{61}v^2}{2v_2}-\lambda_{71}v_2+\frac{\lambda_{71}v_2v^2}{2v_1^2}\right)^2 \ln \frac{m_{H^\pm}^2}{\mu^2} \nonumber\\
 && -6 m_W^2 g_2^2 \left(\frac13 - \ln\frac{m_W^2}{\mu^2}\right) + 
 g_2^4 v_1^2\left(2+3 \ln\frac{m_W^2}{\mu^2}\right) -
 3 m_Z^2 g^2 \left(\frac13 - \ln\frac{m_Z^2}{\mu^2}\right) \nonumber\\
 &&  +\frac{ g^4 v_1^2}{2}\left(2+3 \ln\frac{m_Z^2}{\mu^2}\right) + 
 24 m_b^2 Y_b^2 \left(1 - \ln\frac{m_b^2}{\mu^2}\right) - 24 v_1^2 Y_b^4 \ln \frac{m_b^2}{\mu^2} \nonumber\\
 && + 8 m_{\tau}^2 Y_{\tau}^2 \left(1 - \ln\frac{m_{\tau}^2}{\mu^2}\right) - 
 8 v_1^2 Y_{\tau}^4 \ln \frac{m_{\tau}^2}{\mu^2} + 
 8 \left[ F(h,1)\right]^2 
 \ln\frac{m_h^2}{\mu^2} \nonumber\\
 && \left. +4 m_h^2 G(h,1) \left(\ln\frac{m_h^2}{\mu^2} -1 \right)
 + 8 \left[ F(H,1)\right]^2 \ln\frac{m_H^2}{\mu^2}
 +4 m_H^2 G(H,1) \left(\ln\frac{m_H^2}{\mu^2} -1 \right)
 \right] \,, \nonumber\\
\left.\frac{\partial^2 V_1}{\partial \phi_2^2}\right\vert_{\phi_{c2}=v_2}&=& 
\frac{1}{64\pi^2} \left[ 4 m_A^2 \left(\frac{m_{12}^{2R} v_1}{v_2^3} - \lambda_{51}-\frac{3\lambda_{71}v_2}{2v_1}
+\frac{\lambda_{61}v_1}{2v_2}
-\frac{\lambda_{61}v_1v^2}{2v_2^3}\right)\left(\ln \frac{m_A^2}{\mu^2} - 1\right) \right. \nonumber\\
&&+ 8 \left(\frac{m_{12}^{2R}}{2v_1}\left(1-\frac{v_1^2}{v_2^2}\right)-v_2 \lambda_{51}-\frac{\lambda_{71}v^2}{4v_1}-\frac{\lambda_{71}v_2^2}{2v_1}-
\frac{\lambda_{61}v_1}{2}+\frac{\lambda_{61}v_1v^2}{4v_2^2}\right)^2 \ln \frac{m_A^2}{\mu^2} \nonumber\\
 && + 8 m_{H^\pm}^2 \left(\frac{m_{12}^{2R} v_1}{v_2^3} - \frac12\lambda_{45}-\frac{3\lambda_{71}v_2}{2v_1}
+\frac{\lambda_{61}v_1}{2v_2}
-\frac{\lambda_{61}v_1v^2}{2v_2^3}\right)\left(\ln \frac{m_{H^\pm}^2}{\mu^2} - 1\right) \nonumber\\
&&+ 4  \left(\frac{m_{12}^{2R}}{v_1}\left(1-\frac{v_1^2}{v_2^2}\right)-v_2 \lambda_{45}-\frac{\lambda_{71}v_2^2}{v_1}-\frac{\lambda_{71}v^2}{2v_1}-\lambda_{61}v_1+\frac{\lambda_{61}v_1v^2}{2v_2^2}\right)^2 \ln \frac{m_{H^\pm}^2}{\mu^2} \nonumber\\
 && -6 m_W^2 g_2^2 \left(\frac13 - \ln\frac{m_W^2}{\mu^2}\right) + 
 g_2^4 v_2^2\left(2+3 \ln\frac{m_W^2}{\mu^2}\right) -3 m_Z^2 g^2 \left(\frac13 - 
 \ln\frac{m_Z^2}{\mu^2}\right) \nonumber\\
 &&  +\frac{ g^4 v_2^2}{2}\left(2+3 \ln\frac{m_Z^2}{\mu^2}\right) + 
 24 m_t^2 Y_t^2 \left(1 - \ln\frac{m_t^2}{\mu^2}\right) - 24 v_2^2 Y_t^4 \ln \frac{m_t^2}{\mu^2} \nonumber\\
 && +4 m_h^2 G(h,2) \left(\ln\frac{m_h^2}{\mu^2} -1 \right)
 + 8 \left[ F(h,2)\right]^2 \ln\frac{m_h^2}{\mu^2}
 +4 m_H^2 G(H,2) \left(\ln\frac{m_H^2}{\mu^2} -1 \right)\nonumber\\
 && \left. + 8 \left[F(H,2)\right]^2 \ln\frac{m_H^2}{\mu^2} \right]\,. \nonumber\\
 \eea
 
 Here,
 \bea
 G(h,1)
 &=& \frac14\left(2\lambda_1 - \frac{2m_{12}^{2R}}{v_1 v_2} 
 +\frac{2m_{12}^{2R} v^2}{v_1^3 v_2}-\frac{\lambda_{71}v_2^3}{v_1^3}-\frac{3\lambda_{61}v_1}{v_2}\right)\nonumber\\
 && -\frac{1}{4\sqrt{a}} \left[ \left( \lambda_1v_1^2 - \lambda_2 v_2^2 -\frac{m_{12}^{2R}}{v_1 v_2} 
 \left(v_1^2 - v_2^2\right)+\frac{3v_1v_2\left(\lambda_{61} -\lambda_{71}\right)}{2}
  -\frac{\lambda_{71}v_2^3}{2v_1}
 +\frac{\lambda_{61}v_1^3}{2v_2} \right)\right. \nonumber\\
 &&\times \left(\lambda_1 + \frac{m_{12}^{2R}}{v_1 v_2} 
 -\frac{m_{12}^{2R}}{v_1^3 v_2}\left(v_1^2-v_2^2\right)-\frac{\lambda_{71}v_2^3}{2v_1^3}+\frac{3\lambda_{61}v_1}{2v_2}\right) \nonumber\\
 && +2\, \left(\lambda_1 v_1 -\frac{m_{12}^{2R}}{v_2} + \frac{m_{12}^{2R}}{2v_1^2 v_2}\left(v_1^2-v_2^2\right)+\frac{3v_2}{4}\left(\lambda_{61}-\lambda_{71}\right)+\frac{\lambda_{71}v_2^3}{4v_1^2}
 +\frac{3\lambda_{61}v_1^2}{4v_2}\right)^2 \nonumber\\
 &&\left. +2 \left(\lambda_{345} v_2+3\lambda_{61}v_1\right)^2+6\lambda_{61}\left(-m_{12}^{2R} + 
 \lambda_{345}v_1v_2+\frac{3\lambda_{61}v_1^2}{2}+\frac{3\lambda_{71}v_2^2}{2}\right)\right] \nonumber\\
 && + \frac{1}{8a^{3/2}} 
 \left[ \left( \lambda_1 v_1^2 - \lambda_2 v_2^2 -\frac{m_{12}^{2R}}{v_1 v_2} \left(v_1^2-v_2^2\right)
 +\frac{3v_1v_2\left(\lambda_{61}-\lambda_{71}\right)}{2}
 -\frac{\lambda_{71}v_2^3}{2v_1}+\frac{\lambda_{61}v_1^3}{2v_2}\right) \right.\nonumber\\
 &&\times\left(\lambda_1 v_1 -\frac{m_{12}^{2R}}{v_2}
 + \left(v_1^2- v_2^2\right)\frac{m_{12}^{2R}}{2 v_1^2 v_2}+\frac{3v_2\left(\lambda_{61}-\lambda_{71}\right)}{4}+\frac{\lambda_{71}v_2^3}{4v_1^2}+\frac{3\lambda_{61}v_1^2}{4v_2}\right) \nonumber\\
 && \left. +2\left(-m_{12}^{2R} + \lambda_{345} v_1 v_2+\frac{3\lambda_{61}v_1^2}{2}+
 \frac{3\lambda_{71}v_2^2}{2}\right)\times \left(\lambda_{345} v_2+3\lambda_{61}v_1\right)\right]^2\,, \nonumber\\
\eea
\bea
 G(h,2) 
 &=& \frac14\left(2\lambda_2 - \frac{2m_{12}^{2R}}{v_1v_2} 
 + \frac{2m_{12}^{2R} v^2}{v_2^3 v_1}-\frac{3\lambda_{71}v_2}{v_1}-\frac{\lambda_{61}v_1^3}{v_2^3}\right)\nonumber\\
 && -\frac{1}{4\sqrt{a}} \left[ \left( 
 \lambda_1v_1^2 - \lambda_2 v_2^2 -\frac{m_{12}^{2R}}{v_1 v_2} \left(v_1^2 - v_2^2\right)+\frac{3v_1v_2}{2}\left(\lambda_{61}-\lambda_{71}\right)- 
 \frac{\lambda_{71}v_2^3}{2v_1}+\frac{\lambda_{61}v_1^3}{2v_2} \right) \right. \nonumber\\
 &&\times \left(-\lambda_2 -\frac{m_{12}^{2R}}{v_1 v_2} 
 -\frac{m_{12}^{2R}}{ v_2^3 v_1}\left(v_1^2-v_2^2\right)-\frac{3\lambda_{71}v_2}{2v_1}+\frac{\lambda_{61}v_1^3}{2v_2^3}\right) \nonumber\\
 && +2\, \left( -\lambda_2 v_2 +\frac{m_{12}^{2R}}{v_1} + \frac{m_{12}^{2R}}{2v_2^2 v_1}\left(v_1^2-v_2^2\right)+\frac{3v_1 \left(\lambda_{61}-\lambda_{71}\right)}{4} 
 -\frac{3\lambda_{71}v_2^2}{4v_1}-\frac{\lambda_{61}v_1^3}{4v_2^2} \right)^2 \nonumber\\
 &&\left. +6\lambda_{71}\left(-m_{12}^{2R}+\lambda_{345}v_1v_2+\frac{3\lambda_{61}v_1^2}{2}+\frac{3\lambda_{71}v_2^2}{2}\right)+2\left(\lambda_{345}v_1+3\lambda_{71}v_2\right)^2 \right] \nonumber\\
 && + \frac{1}{8a^{3/2}}\left[ \left( 
 \lambda_1 v_1^2 - \lambda_2 v_2^2 -\frac{m_{12}^{2R}}{v_1 v_2} \left(v_1^2-v_2^2\right)+\frac{3v_1v_2\left(\lambda_{61}-\lambda_{71}\right)}{2} 
 -\frac{\lambda_{71}v_2^3}{2v_1}+\frac{\lambda_{61}v_1^3}{2v_2}\right) \right. \times\nonumber\\
 &&\times\left(-\lambda_2 v_2 +\frac{m_{12}^{2R}}{v_1}
 + \left(v_1^2- v_2^2\right)\frac{m_{12}^{2R}}{2 v_2^2 v_1}+\frac{3v_1\left(\lambda_{61}-\lambda_{71}\right)}{4}-\frac{3\lambda_{71}v_2^2}{4v_1}-\frac{\lambda_{61}v_1^3}{4v_2^2}\right) \nonumber\\
 &&\left. +2\left(-m_{12}^{2R} + \lambda_{345} v_1 v_2+\frac{3\lambda_{61}v_1^2}{2}+
 \frac{3\lambda_{71}v_2^2}{2}\right)\times\left(\lambda_{345} v_1+3\lambda_{71}v_2\right)\right]^2\,,
 \eea
\bea
G(H,1)&=& \frac14\left(2\lambda_1 - \frac{2m_{12}^{2R}}{v_1 v_2} 
 +\frac{2m_{12}^{2R} v^2}{v_1^3 v_2}-\frac{\lambda_{71}v_2^3}{v_1^3}-\frac{3\lambda_{61}v_1}{v_2}\right)\nonumber\\
 && +\frac{1}{4\sqrt{a}} \left[ \left( 
 \lambda_1v_1^2 - \lambda_2 v_2^2 -\frac{m_{12}^{2R}}{v_1 v_2} \left(v_1^2 - 
 v_2^2\right)+\frac{3v_1v_2\left(\lambda_{61}-\lambda_{71}\right)}{2} 
  -\frac{\lambda_{71}v_2^3}{2v_1}
 +\frac{\lambda_{61}v_1^3}{2v_2}\right)\right. \nonumber\\
 &&\times \left(\lambda_1 + \frac{m_{12}^{2R}}{v_1 v_2} 
 -\frac{m_{12}^{2R}}{v_1^3 v_2}\left(v_1^2-v_2^2\right)-\frac{\lambda_{71}v_2^3}{2v_1^3}+\frac{3\lambda_{61}v_1}{2v_2}\right) \nonumber\\
 && +2 \left( \lambda_1 v_1 -\frac{m_{12}^{2R}}{v_2} + 
 \frac{m_{12}^{2R}}{2v_1^2 v_2}\left(v_1^2-v_2^2\right)+
 \frac{3v_2}{4}\left(\lambda_{61}-\lambda_{71}\right) 
  +\frac{\lambda_{71}v_2^3}{4v_1^2}
 +\frac{3\lambda_{61}v_1^2}{4v_2}\right)^2 \nonumber\\
 && \left. 
 +2 \left(\lambda_{345} v_2+3\lambda_{61}v_1\right)^2+
 6\lambda_{61}\left(-m_{12}^{2R}+\lambda_{345}v_1v_2+\frac{3\lambda_{61}v_1^2}{2}+\frac{3\lambda_{71}v_2^2}{2}\right)\right] \nonumber\\
 && - \frac{1}{8a^{3/2}}
 \left[ \left(
 \lambda_1 v_1^2 - \lambda_2 v_2^2 -\frac{m_{12}^{2R}}{v_1 v_2} \left(v_1^2-v_2^2\right)+
 \frac{3v_1v_2\left(\lambda_{61}-\lambda_{71}\right)}{2}
 -\frac{\lambda_{71}v_2^3}{2v_1}+\frac{\lambda_{61}v_1^3}{2v_2}\right) \right.\nonumber\\
 &&\times\left(\lambda_1 v_1 -\frac{m_{12}^{2R}}{v_2}
 + \left(v_1^2- v_2^2\right)\frac{m_{12}^{2R}}{2 v_1^2 v_2}+\frac{3v_2\left(\lambda_{61}-
 \lambda_{71}\right)}{4}+\frac{\lambda_{71}v_2^3}{4v_1^2}+\frac{3\lambda_{61}v_1^2}{4v_2}\right) \nonumber\\
 && +2\left. \left(-m_{12}^{2R} + \lambda_{345} v_1 v_2+\frac{3\lambda_{61}v_1^2}{2}+
 \frac{3\lambda_{71}v_2^2}{2}\right)\times \left(\lambda_{345} v_2+3\lambda_{61}v_1\right)\right]^2\,, \nonumber\\
\eea
 \bea
 G(H,2)
 &=& \frac14\left(2\lambda_2 - \frac{2m_{12}^{2R}}{v_1v_2} 
 + \frac{2m_{12}^{2R} v^2}{v_2^3 v_1}-\frac{3\lambda_{71}v_2}{v_1}-\frac{\lambda_{61}v_1^3}{v_2^3}\right)\nonumber\\
 && +\frac{1}{4\sqrt{a}} \left[ \left( 
 \lambda_1v_1^2 - \lambda_2 v_2^2 -\frac{m_{12}^{2R}}{v_1 v_2} \left(v_1^2 - v_2^2\right)+\frac{3v_1v_2}{2}\left(\lambda_{61}-\lambda_{71}\right) 
  -\frac{\lambda_{71}v_2^3}{2v_1}+\frac{\lambda_{61}v_1^3}{2v_2} \right)\right. \nonumber\\
 &&\times \left(-\lambda_2 -\frac{m_{12}^{2R}}{v_1 v_2} 
 -\frac{m_{12}^{2R}}{ v_2^3 v_1}\left(v_1^2-v_2^2\right)-\frac{3\lambda_{71}v_2}{2v_1}+\frac{\lambda_{61}v_1^3}{2v_2^3}\right) \nonumber\\
 && +2\, \left( -\lambda_2 v_2 +\frac{m_{12}^{2R}}{v_1} + \frac{m_{12}^{2R}}{2v_2^2 v_1}\left(v_1^2-v_2^2\right)+\frac{3v_1 \left(\lambda_{61}-\lambda_{71}\right)}{4} 
 -\frac{3\lambda_{71}v_2^2}{4v_1}-\frac{\lambda_{61}v_1^3}{4v_2^2} \right)^2 \nonumber\\
 && \left. +6\lambda_{71}\left(-m_{12}^{2R}+\lambda_{345}v_1v_2+\frac{3\lambda_{61}v_1^2}{2}+\frac{3\lambda_{71}v_2^2}{2}\right)+2\left(\lambda_{345}v_1+3\lambda_{71}v_2\right)^2 \right] \nonumber\\
 && - \frac{1}{8a^{3/2}} \left[\left( 
 \lambda_1 v_1^2 - \lambda_2 v_2^2 -\frac{m_{12}^{2R}}{v_1 v_2} \left(v_1^2-v_2^2\right)+\frac{3v_1v_2\left(\lambda_{61}-\lambda_{71}\right)}{2} 
 -\frac{\lambda_{71}v_2^3}{2v_1}+\frac{\lambda_{61}v_1^3}{2v_2} \right)\right. \times\nonumber\\
 &&\times\left(-\lambda_2 v_2 +\frac{m_{12}^{2R}}{v_1}
 + \left(v_1^2- v_2^2\right)\frac{m_{12}^{2R}}{2 v_2^2 v_1}+\frac{3v_1\left(\lambda_{61}-\lambda_{71}\right)}{4}-\frac{3\lambda_{71}v_2^2}{4v_1}-\frac{\lambda_{61}v_1^3}{4v_2^2}\right) \nonumber\\
 && \left. +2\left(-m_{12}^{2R} + \lambda_{345} v_1 v_2+\frac{3\lambda_{61}v_1^2}{2}+\frac{3\lambda_{71}v_2^2}{2}\right)\times\left(\lambda_{345} v_1+3\lambda_{71}v_2\right) \right]^2\,. \nonumber\\
 \eea
\end{appendix}

\end{document}